\documentclass[10pt,a4paper,twocolumn]{IEEEtran}
\usepackage{cite}
\usepackage{amsmath,amssymb,amsfonts}
\usepackage{algorithmic}
\usepackage{graphicx}
\usepackage{textcomp}
\def\BibTeX{{\rm B\kern-.05em{\sc i\kern-.025em b}\kern-.08em
    T\kern-.1667em\lower.7ex\hbox{E}\kern-.125emX}}
\begin{document}
\setlength{\topsep}{0.0cm}
\setlength{\itemsep}{0.0em}
\title{{\large Interaction Identification of a Heterogeneous NDS with Quadratic-Bilinear Subsystems}}
\vspace{-0.5cm}
\author{{\small Tong Zhou and Yubing Li}
\thanks{ This work was supported in part by the NNSFC, China under Grant
	 62373212, 62127809, 61733008, and 52061635102, and by the BNRist project under Grant BNR2024TD03003. }
\thanks{T. Zhou is with
the Department of Automation and BNRist, Tsinghua University, Beijing, 100084, China (e-mail: tzhou@mail.tsinghua.edu.cn).}
\thanks{Y. Li is with
the Department of Automation, Tsinghua University, Beijing, 100084, China (e-mail: liyb20@mails.tsinghua.edu.cn).}
\vspace*{-1.0cm}}

\maketitle
\vspace{-0.5cm}
\begin{abstract}
This paper attacks time-domain identification for interaction parameters of a heterogeneous networked dynamic system (NDS), with each of its subsystems being described by a continuous-time descriptor quadratic-bilinear time-invariant (QBTI) model. The obtained results can also be applied to parameter estimations for a lumped QBTI system. No restrictions are put on the sampling rate. Explicit formulas are derived respectively for the transient and steady-state responses of the NDS, provided that the probing signal is generated by a linear time invariant (LTI) system. Some relations have been derived between the NDS steady-state response and its frequency domain input-output mappings. These relations reveal that the value of some NDS associated generalized TFMs can in principle be estimated at almost any interested point of the imaginary axis from time-domain input-output experimental data, as well as its derivatives and a right tangential interpolation along an arbitrary  direction. Based on these relations, an estimation algorithm is suggested respectively for the parameters of the NDS and the values of these generalized TFMs. A numerical example is included to illustrate characteristics of the suggested estimation algorithms.
\end{abstract}


\addtolength{\abovedisplayskip}{-0.10cm}
\addtolength{\belowdisplayskip}{-0.10cm}

\addtolength{\abovedisplayshortskip}{-0.10cm}
\addtolength{\belowdisplayshortskip}{-0.10cm}

\setlength{\footskip}{-1.0cm}
\addtolength{\footskip}{-1.20cm}

\vspace*{-0.5cm}
\section{Introduction}
\label{sec:introduction}

\setlength{\itemsep}{0.0em}
In various fields such as engineering, biology, etc., there exist systems that are constituted from numerous subsystems. Revealing the dynamics and structure of these systems from experimental data are essential from many aspects of applications, such as data analysis and processing, system analysis and design, etc. \cite{cw2020,Fortunato2010,zg2012,dtrf2019,vtc2021,ptt2019,wvd2018,zly2024,Zhou2022,zy2022}.

In describing the dynamics of a nonlinear plant, a quadratic-bilinear (QB) model is extensively utilized \cite{Faruqi2005}. It is also well known that through the McCormick relaxation, which calculates derivatives of a function and/or adds some algebraic equations, several types of smooth analytic nonlinearities can be transformed into a quadratic-bilinear form \cite{abg2020,Gosea2022}. While there are various nonlinear dynamic systems that can be directly described by or transformed into a QB model, researches are mainly focused on model reduction. To be more specific, limited to our knowledge, there are still no studies attacking parameter estimations even for a quadratic-bilinear time-invariant (QBTI) system \cite{kga2022}.

In this paper, we investigates estimation of  subsystem interactions for a networked dynamic system (NDS), in which the dynamics of each subsystem is represented by a QBTI model. It is not required that every subsystem has the same dynamics, and interactions among subsystem are only asked to be linear. It is proved that under such a situation, the assembly system can still be described by a QBTI model, with its system matrices being a linear fractional transform (LFT) of its subsystem connection matrix (SCM) or its subsystem interaction parameters (SIP). This makes the obtained results also applicable to parameter estimations for a lumped QBTI system.

On the other hand, when a QBTI system is stimulated by the outputs of a linear time invariant (LTI) system, an explicit formula is derived respectively for the transient response and the steady-state response of the QBTI system, which extends results of a linear NDS given in \cite{Zhou2025}. It has been made clear that this steady-state response depends linearly on the values of the transfer function matrix (TFM) of the linear part of the QBTI system and some of its multi-dimensional or generalized TFMs at some particular locations. These TFMs are completely determined by the system matrices of the QBTI system, while the locations are given by a linear combination of the eigenvalues of the probing signal generation system (PSGS) with some nonnegative integer coefficients. Different from a linear plant, in the steady-state response of a QBTI system, not only each of the dynamic modes of the PSGS, but also their linear combinations with nonnegative integral coefficients, are included. Moreover, these formulas also reveal that in the transient response of the QBTI system, in addition to the dynamic modes of its linear part, their linear combinations with the dynamic modes of the PSGS, are also available. The existence of these linear mode combinations makes the associated subsystem interaction estimation problem mathematically more involved, and restricts selections of sampling instants and locations for nonparametric estimations, etc. As a byproduct, these expressions provides analytic expressions for harmonics in the time-domain response of a nonlinear dynamic system through its frequency-domain responses, and therefore establish some relations between its time and frequency-domain characteristics.

On the basis of this expression for the steady-state response of the QBTI system, as well as orthogonality properties of sinusoidal signals, an estimate is obtained for a tangential interpolation of the aforementioned TFMs. Different from an NDS consisted of LTI subsystems, non-uniform sampling is no longer permitted, but the sampling rate is still allowed to be slower than the Nyquist frequency of the linear part. In addition, it has also been shown that the aforementioned TFM and generalized TFMs can be expressed through an LFT of the SCM/SIP vector of the NDS. From these expressions, an estimate is derived for the SIPs of the NDS. These results can be directly applied to parameter estimations for a lumped QBTI system, as long as the TFM of its linear part and/or some of its generalized TFMs depend on the parameters to be estimated through an LFT.

The remaining of this paper is organized as follows. At first, in Section 2, problem descriptions and some preliminary results are given. Decompositions of time-domain system response are attacked in Section 3. Section 4 investigates nonparametric and parametric estimations for the NDS. A numerical example is reported in Section 5, demonstrating properties of the suggested estimation algorithms. Some concluding remarks are given in Section 6 in which several further issues are discussed for NDS structure identification. Finally, an appendix is included to give proofs of some technical results.

The following notation and symbols are adopted in this paper. $ \mathbb{R}^{n}/\mathbb{C}^{n}$ ($\mathbb{R}^{m\times n}/\mathbb{C}^{m\times n}$) represents respectively the set of $ n $ ($ m\times n $) dimensional real/complex vectors (matrices), while $||\cdot||$ the Euclidean norm of a vector or its induced norm of a matrix. $\overline{\sigma}(\star)$ stands for the maximum singular value of a matrix, while $ \star^{T} $ its transpose. For a column (row) vector/matrix $\star$, $[\star]_{i}$ denotes its $i$-th row (column) elemnt/column vector. ${\it diag}\!\{\star_{i}\:|^n_{i=1}\}$ and $ {\it  col}\!\left\lbrace \star_{i}\:|_{i=1}^{n} \right\rbrace  $ represent  the matrices composed of $\star _{i}|_{i=1}^{n}$ stacking respectively diagonally and vertically. For a complex variable/vector/matrix, the superscript $*$, $[r]$ and $[i]$ denote respectively its conjugate, real part and imaginary part. $I_{n} $ stands for the $ n $ dimensional identity matrix, while $ 0_{m\times n} $ the $ m\times n $ dimensional zero matrix. When the dimension is obvious or insignificant, these two matrices are abbreviated as $ I $ and $ 0 $, respectively.  ${\cal L}(\cdot)$ is adopted to denote the Laplace transformation of a vector valued function (VVF) of time, while ${\cal L}^{-1}(\cdot)$ its inverse transformation. The imaginary unit  $\sqrt{-1}$ is denoted by $\mathbf{i}$. With a little abuse of notations, the value of the summation $\sum_{k=k_{low}}^{k_{up}}f(k)$ with $k_{up} < k_{low}$ is defined to be zero.

\vspace{-0.25cm}
\section{Problem Formulation and Preliminaries}\label{sec:problem}

\setcounter{enumi}{\value{equation}}
\addtocounter{enumi}{1}
\setcounter{equation}{0}
\renewcommand{\theequation}{\theenumi\alph{equation}}

Consider a continuous-time NDS $\mathbf{\Sigma}_{p}$ with its subsystems having nonlinear dynamics that may be distinctive from each other and can be described by some QBTI models, and their direct interactions are linear but otherwise arbitrary. More precisely, assume that the NDS $\mathbf{\Sigma}_{p}$ consists of $N_{p}$ subsystems, with the dynamics of its $i$-th subsystem $\mathbf{\Sigma}_{p,i}$, $i=1,2,\cdots,N_{p}$, being described by the following 3 equations,
\begin{eqnarray}
& & \hspace*{-0.8cm} E(i)\dot{x}(t,i)\!=\! A_{xx}(i)x(t,i) \!+\! B_{xv}(i)v(t,i) \!+\! B_{xu}(i)u(t,i) \!+ \nonumber \\
           & & \hspace*{1.4cm}  \Gamma_{xx}(i)\left(x(t,i) \otimes x(t,i)\right) \!+\! \Gamma_{xv}(i)\left(x(t,i) \otimes \right. \nonumber \\
           & & \hspace*{1.4cm} \left.v(t,i)\right)
                    \!+\! \Gamma_{xu}(i)\left(x(t,i) \otimes u(t,i)\right)
\label{subsystem-1} \\
& & \hspace*{-0.8cm} z(t,i) \!=\! C_{zx}(i)x(t,i) \!+\! D_{zv}(i)v(t,i) \!+\!  D_{zu}(i) u(t,i)
\label{subsystem-2} \\
& & \hspace*{-0.8cm}y(t,i) \!=\! C_{yx}(i)x(t,i) \!+\! D_{yv}(i)v(t,i) \!+\!  D_{yu}(i) u(t,i)
\label{subsystem-3}
\end{eqnarray}
in which $E(i)$ is a real square matrix that may not be invertible. In actual applications, this matrix is usually utilized to reflect constraints on system variables, etc. $t$ stands for the temporal variable, while $x(t,i)$ the state vector of the $i$-th subsystem $\mathbf{\Sigma}_{p,i}$. Outputs and inputs of this subsystem are divided into internal and external parts, in which the internal ones are used to represent subsystem interactions, while the external ones are actual NDS inputs or outputs. In particular,  $u(t,i)$ and $y(t,i)$ are used to denote respectively the external input/output vectors of Subsystem $\mathbf{\Sigma}_{p,i}$, while $v(t,i)$ and $z(t,i)$ respectively its internal input/output vectors, meaning signals obtained from other subsystems and signals sent to other subsystems.

\setcounter{equation}{\value{enumi}}
\renewcommand{\theequation}{\arabic{equation}}
In addition, subsystem interactions of the whole NDS $\mathbf{\Sigma}_{p}$ are described by the following equation,
\begin{equation}
\hspace*{-0.25cm} v(t)\!=\!\Theta(\theta) z(t) \hspace{0.15cm} {\rm with} \hspace{0.15cm} \Theta(\theta)\!\!=\!\!\sum\limits_{i=1}^{m_{\theta}}\theta_{i}\Theta_{i},
\hspace{0.10cm}  \theta \!=\!
\left[\theta_{1}\; \cdots\; \theta_{m_{\theta}}\right]^{T}
\label{scm}
\end{equation}
in which $v(t)$ and $z(t)$ are defined respectively as $z(t)=col\{z(t,i)|^{N_{p}}_{i=1}\}$ and $v(t)=col\{v(t,i)|^{N_{p}}_{i=1}\}$, being  assembly expressions for all the internal input and output vectors of the NDS $\mathbf{\Sigma}_{p}$. Matrix $\Theta(\theta)$ depicts interactions among NDS subsystems, in which $\theta_{i}|_{i=1}^{m_{\theta}}$ are parameters to be estimated, while $\Theta_{i}|_{i=1}^{m_{\theta}}$ are some known real matrices reflecting available information about the topology of the NDS $\mathbf{\Sigma}_{p}$ gained from its working principles, etc. Note that when each subsystem $\mathbf{\Sigma}_{p,i}$ and each nonzero element of the matrix $\Theta(\theta)$ are respectively considered as a node and a directed weighted edge, a graph can be constructed for the NDS $\mathbf{\Sigma}_{p}$, which is known as its topology or structure, representing direct interactions among NDS subsystems. In addition, the matrix $\Theta(\theta)$ is usually called subsystem connection matrix, while $\theta_{i}|_{i=1}^{m_{\theta}}$ subsystem interaction parameters \cite{cw2020,ptt2019,Zhou2022}.

It is worthwhile to mention that in general, there are repeated elements in the vector $x(t,i) \otimes x(t,i)$ for each $i=1,2,\cdots,N_{p}$. This repetition may make some parameters in the SCM $\Theta(\theta)$ unidentifiable. To avoid occurrence of this problem, it is assumed without any loss of generality that in all the columns of the matrix $\Gamma_{xx}(i)$, that are associated with the same elements of the vector $x(t,i) \otimes x(t,i)$, except the first column from its left, all the other columns are set to be a column with all elements being zero.

Throughout this paper, the dimension of a vector $\star(t,i)$ with $i=1,2,\cdots,N_{p}$ and $\star$ being $u$, $v$, $x$, $y$ or $z$, is denoted by   $m_{\star,i}$. Using these symbols, define an integer $m_\star$ as $m_\star=\sum\nolimits_{i=1}^{N_{p}} m_{\star,i}$. Then the SCM $\Theta(\theta)$ is clearly a $m_v\times m_z$ dimensional real matrix. Moreover, denote vectors $col\{x(t,i)|^{N_{p}}_{i=1}\}$, $col\{u(t,i)|^{N_{p}}_{i=1}\}$ and $col\{y(t,i)|^{N_{p}}_{i=1}\}$ respectively by $x(t)$, $u(t)$ and $y(t)$. To clarify that both the NDS $\mathbf{\Sigma}_{p}$ and its external output vector $y(t)$ depend on its SIP vector $\theta$, they are sometimes also written respectively as $\mathbf{\Sigma}_{p}(\theta)$ and $y(t,\theta)$. This expression is adopted also for other symbols.

On the other hand, assume that the external input signal vector $u(t,i)$ of the $i$-th subsystem $\mathbf{\Sigma}_{p,i}$ is generated by the following autonomous LTI system $\mathbf{\Sigma}_{s,i}$, with its state vector $\xi(t,i)$ belonging to $\mathbb{R}^{m_{\xi,i}}$ and its system matrices $\Xi(i)$ and $\Pi(i)$ having compatible dimensions, that is $\Xi(i) \in \mathbb{R}^{m_{\xi,i}\times m_{\xi,i}}$ and $\Pi \in \mathbb{R}^{m_{y,i}\times m_{\xi,i}}$,
\begin{equation}
	\dot{\xi}(t,i)  = \Xi(i) \xi(t,i), \hspace{0.5cm}
	u(t,i) = \Pi(i) \xi(t,i)  \label{signal-1}
\end{equation}

The objectives of this paper are to develop an estimation procedure for the SIP vector $\theta$, using measured values of the external output vector $y(t)$ of the NDS $\mathbf{\Sigma}_{p}$ at some uniformly distributed sampling instants, denote them by $t_{k}=kT$, $k=1,2,\cdots,N_{d}$, under the condition that for each $i=1,2,\cdots, N_{p}$, all the system matrices of the PSGS $\mathbf{\Sigma}_{s,i}$, as well as its initial condition $\xi(0,i)$, are exactly known. Here, $T$ stands for a sampling period.

\newtheorem{Assumption}{Assumption}

\newtheorem{Lemma}{Lemma}
The next results decompose the state vector of a QBTI system into the state vectors of an infinite series of LTI systems, that are closely related to the extensively known Volterra series representation of a nonlinear dynamic system \cite{abg2020,Gosea2022}.

\begin{Lemma}
For each admissible pair of initial conditions and input signal, the solution to the following QBTI system
\begin{displaymath}
\hspace*{-0.2cm} E \dot{x}(t) \!=\! Ax(t) \!+\! Bu(t) \!+\! \Gamma_{x} \!\left(x(t) \!\otimes\! x(t)\right) \!+\! \Gamma_{u} \!\left(x(t) \!\otimes\! u(t)\right)
\end{displaymath}
can be equivalently written as
\begin{displaymath}
x(t)=\sum_{k=1}^{\infty}x_{k}(t) \hspace{0.25cm}{\rm with} \hspace{0.25cm} E \dot{x}_{1}(t) \!=\! Ax_{1}(t)+Bu(t)
\end{displaymath}
and for each $k \geq 2$, $x_{k}(t)$ is the solution to the following LTI system
\begin{displaymath}
\hspace*{-0.2cm}
E \dot{x}_{k}(t) \!=\! Ax_{k}(t) \!+\! \Gamma_{x} \!\!\sum_{l=1}^{k \!-\!1}\!\!\left(x_{l}(t) \!\otimes\! x_{k-l}(t)\right) \!+\!
           \Gamma_{u}\!\left(x_{k \!-\! 1}(t) \!\otimes\! u(t)\right)
\end{displaymath}
\label{lemma:2}
\end{Lemma}

In system analysis and synthesis, the $x_{1}(t)$ related dynamics in the above decomposition for $x(t)$ is usually called as the linear part of the QBTI system \cite{abg2020,Gosea2022}.

When a regular descriptor system is stimulated by the output of an LTI system, explicit expressions can be obtained for its time-domain responses that establish some simple and analytic relations between its time-domain and frequency-domain characteristics \cite{Zhou2025}. More precisely, consider the following LTI continuous-time descriptor system $\mathbf{\Sigma}_{d}$ with an initial state vector $x(0)\in \mathbb{R}^{m_{x}}$,
\begin{equation}
E \dot{x}(t)  \!=\! Ax(t) \!+\! Bu(t), \hspace{0.50cm}
y(t) \!=\! Cx(t) \!+\! Du(t)
\label{eqn:plant-1}
\end{equation}
Assume that it is regular and stimulated by the output of the following LTI continuous-time system $\mathbf{\Sigma}_{s}$ with an initial state vector $\xi(0)\in \mathbb{R}^{m_{\xi}}$,
\begin{equation}
	\dot{\xi}(t)  = \Xi \xi(t), \hspace{0.5cm}
	u(t) = \Pi \xi(t)
\label{eqn:signal-1}
\end{equation}
Then the following results can be derived straightforwardly from Lemma 3 and Theorem 1 of \cite{Zhou2025}.

\begin{Lemma}
Assume that all the eigenvalues of System $\mathbf{\Sigma}_{s}$, denote them by $\lambda_{s,i}|_{i=1}^{m_{\xi}}$, are different from each generalized eigenvalue of System $\mathbf{\Sigma}_{d}$. Moreover, assume that there is an invertible matrix $T_{s}\in \mathbb{C}^{m_{\xi}\times m_{\xi}}$, such that $\Xi=T_{s}diag\left\{\lambda_{s,i}|_{i=1}^{m_{\xi}}\right\}T_{s}^{-1}$. For each $i=1,2,\cdots,m_{\xi}$, define a vector $\psi_{u}(i)$ as
\begin{displaymath}
\psi_{u}(i) = \left[T_{s}^{-1}\xi(0)\right]_{\!i} \Pi \left[T_{s}\right]_{i}
\end{displaymath}
Moreover, denote the TFM of System $\mathbf{\Sigma}_{d}$ by $H(s)$. Then there exists a real constant matrix $X\in \mathbb{R}^{m_{x}\times m_{\xi}}$, such that
\begin{eqnarray*}
& & x(t) \!=\! {\cal L}^{-1}\!\!\left\{(sE \!-\! A)^{-1}\right\}E\left[x(0) \!-\! X\xi(0)\right] \!+\!  x_{s}(t) \\
& & y(t) \!=\! C{\cal L}^{-1}\!\!\left\{(sE \!-\! A)^{-1}\right\}E\left[x(0) \!-\! X\xi(0)\right] + y_{s}(t)
\end{eqnarray*}
in which
\begin{eqnarray*}
& &\hspace*{-0.8cm}
x_{s}(t) \!=\! \sum_{i=1}^{m_{\xi}}\!\!\left\{\!e^{\lambda_{s,i}t} \!\!\times\!\! \left[(\lambda_{s,i}E-A)^{-1} \!B\right] \!\!\times\!\! \psi_{u}(i)\!\right\}  \\
& & \hspace*{-0.8cm}
y_{s}(t) \!=\! \sum_{i=1}^{m_{\xi}}\left\{e^{\lambda_{s,i}t}\times H(\lambda_{s,i}) \times \psi_{u}(i)\right\}  	
\end{eqnarray*}
\label{lemma:3}
\end{Lemma}

It is worthwhile to mention that while $\lambda_{s,i}$ and $\psi_{u}(i)$, $i=1,2,\cdots,m_{\xi}$, may take a complex value, both $x_{s}(t)$ and $y_{s}(t)$ are real valued. This is always guaranteed by that all the system matrices of Systems $\mathbf{\Sigma}_{d}$ and $\mathbf{\Sigma}_{s}$ are real valued \cite{Zhou2025}.

While results are also available in \cite{Zhou2025} even when the system matrix $\Xi$ is not similar to a diagonal matrix, this case is not discussed in this paper for avoiding an awkward presentation. But it is worthwhile to mention that the results of this paper can be extended directly to that case by the same token of \cite{Zhou2025}.

\vspace{-0.25cm}
\section{System output decomposition}\label{sec:res-dec}

To develop an estimation procedure for the SIP vector $\theta$ of the NDS $ \mathbf{\Sigma}_{p} $, a decomposition is derived for its output vector $y(t)$ in this section, which clarifies relations between its steady-state responses and its frequency domain input-output mappings.

To make mathematical derivations more concise, the following matrices are defined. $E=diag\{E(i)|^{N_{p}}_{i=1}\}$, $A_{xx}=diag\{A_{xx}(i)|^{N_{p}}_{i=1}\}$, $B_{\star\#}=diag\{B_{\star\#}(i)|^{N_{p}}_{i=1}\}$, $C_{\star\#}=diag\{C_{\star\#}(i)|^{N_{p}}_{i=1}\}$, $D_{\star\#}=diag\{D_{\star\#}(i)|^{N_{p}}_{i=1}\}$, in which $\star=x,y$ or $z$, $\#=x,u$ or $v$. In addition, for each $i\in\{1,2,\cdots,N_{p}\}$ and $\star\in\{x,u,v\}$, divide the matrix $\Gamma_{x\star}(i)$ into $m_{x,i}$ column blocks with each block having $m_{\star,i}$ columns, and denote the $j$-th column block by $\Gamma_{x\star}(i,j)$, $j=1,2,\cdots,m_{x,i}$. Define a matrix $\Gamma_{x\star}$ as following
\begin{displaymath}
\Gamma_{x\star}=diag\{\left[0\;\Gamma_{\star}(i,1) \;0 \;\Gamma_{\star}(i,2) \;0 \; \cdots \; \Gamma_{\star}(i,m_{x,i})\;0 \right]|^{N_{p}}_{i=1}\}
\end{displaymath}
in which the zero matrices have different dimensions that can be understood from the relations between $x(t)\otimes \star(t)$ and $x(t,i)\otimes \star(t,i)$.

With these symbols, the following relation is established among the internal output vector $z(t)$ of the NDS $ \mathbf{\Sigma}_{p} $, its state vector $x(t)$, and its external input vector $u(t)$,  through substituting Equation (\ref{scm}) into Equation (\ref{subsystem-2}),
\begin{displaymath}
\left[I_{m_z} \!-\! D_{zv}\Theta(\theta)\right] z(t)=C_{zx}x(t) + D_{zu}u(t)
\end{displaymath}

To guarantee that the NDS $ \mathbf{\Sigma}_{p} $ has an unique response under the stimulation of any admissible external input signals with an arbitrary admissible initial state vector, the following assumption is adopted in this paper, which guarantees that the above equation has an unique solution for the internal output vector $z(t)$.

\begin{Assumption}\label{assum:2}
For each $\theta \!\in\! \mathbf{\Theta}$, the NDS $\mathbf{\Sigma}_{p}$ is well-posed, meaning invertibility of the matrix $I_{m_{z}} \!\!-\!\! D_{zv} \Theta(\theta)$.
\end{Assumption}

Specifically, when Assumption \ref{assum:2} is satisfied, the following expression is valid for the internal output vector $z(t)$ of the NDS $ \mathbf{\Sigma}_{p} $,
\begin{equation}
z(t)=\left[I_{m_z} \!-\! D_{zv}\Theta(\theta)\right]^{-1}\left\{C_{zx}x(t) + D_{zu}u(t) \right\}
\end{equation}

\setcounter{enumi}{\value{equation}}
\addtocounter{enumi}{1}
\setcounter{equation}{0}
\renewcommand{\theequation}{\theenumi\alph{equation}}

On the basis of this relation, the next lumped model can be further obtained for the NDS $ \mathbf{\Sigma}_{p} $ from Equations (\ref{subsystem-1}) and (\ref{subsystem-3}).
\begin{eqnarray}
& &	\hspace*{-1.0cm} E \dot{x}(t) \!=\! A(\theta)x(t)+B(\theta)u(t)+\Gamma_{x}(\theta)\left(x(t) \otimes x(t)\right) +  \nonumber \\
           & & \hspace*{2.2cm} \Gamma_{u}(\theta)\left(x(t) \otimes u(t)\right)
\label{plant-1}\\
& &	\hspace*{-1.0cm} y(t) \!=\! C(\theta)x(t)+D(\theta)u(t)   \label{plant-2}
\end{eqnarray}
\hspace*{-0.00cm}in which
\begin{eqnarray}
& &	\hspace*{-0.7cm}		\left[\!\!\!\begin{array}{cc}
				A(\theta) & B(\theta)\\
				C(\theta) & D(\theta)
		\end{array}\!\!\!\right] \!=\!
		\left[\!\!\!\begin{array}{cc}A_{xx} & B_{xu}\\
				C_{yx} & D_{yu}\\
		\end{array}\!\!\!\right] \!\!+\!\!
        \left[\!\!\begin{array}{c}
				B_{xv}\\
				D_{yv}
        \end{array}\!\!\right]\!
        \left[ I_{m_{v}} \!\!-\!\! \Theta(\theta)D_{zv}\right]^{-\!1}\!\!\times \nonumber\\
& &\hspace*{3.2cm}  \Theta(\theta)
        \left[\!\! \begin{array}{cc}
				C_{zx} & D_{zu}
        \end{array} \!\!\right]
\label{plant-3} \\
& &	\hspace*{-0.70cm}
\left[\!\!\!\begin{array}{c}
				\Gamma_{\! x}(\theta)\\
				\Gamma_{\! u}(\theta)
		\end{array}\!\!\!\right] \!\!=\!\!
\left[\!\!\!\begin{array}{c}
				\Gamma_{\! xx}\\
				\Gamma_{\! xu}
		\end{array}\!\!\!\right] \!+\!
\Gamma_{\! xv}\!\!
\left[\!\!\!\begin{array}{c}
				I \!\otimes\! \left(\!\! \Theta(\theta)\!\left( I_{m_{v}} \!\!-\!\! D_{zv}\Theta(\theta)\right)^{-\!1}\!\right) \!C_{\! zx}\\
				I \!\otimes\! \left(\!\! \Theta(\theta)\!\left( I_{m_{v}} \!\!-\!\! D_{zv}\Theta(\theta)\right)^{-\!1}\!\right)\! D_{\! zu}
		\end{array}\!\!\!\right]
\label{plant-4}
\end{eqnarray}

\setcounter{equation}{\value{enumi}}
\renewcommand{\theequation}{\arabic{equation}}

The above expressions reveal that the NDS $ \mathbf{\Sigma}_{p} $ is still a QBTI system, and its system matrices depend on the SIP vector $\theta$ through some LFTs. The latter is completely the same as that for the NDS model adopted in \cite{zy2022,Zhou2022}, which is  constituted from several LTI subsystems.

In addition to these, define vector $\xi(t)$, as well as matrices $\Xi$ and $\Pi$, respectively as $\xi(t) = col\{\xi(t,i)|^{N_{p}}_{i=1}\}$ and
\begin{displaymath}
\Xi = diag\{\Xi(i)|^{N_{p}}_{i=1}\}, \hspace{0.25cm}
\Pi = diag\{\Pi(i)|^{N_{p}}_{i=1}\}
\end{displaymath}
Then the input signal generated by all the PSGSs for the NDS $ \mathbf{\Sigma}_{p} $, that is, $ \mathbf{\Sigma}_{s,i}|_{i=1}^{N_{p}} $,  can be equivalently written into an assembly form, which is completely the same of System $ \mathbf{\Sigma}_{s}$ described by Equation (\ref{eqn:signal-1}).

From Equations (\ref{plant-1}) and (\ref{eqn:signal-1}), it is clear that properties of the time-domain responses of the NDS $\mathbf{\Sigma}_{p}$ can be analyzed using the decompositions of Lemmas \ref{lemma:2} and \ref{lemma:3}.

To make parameter estimations meaningful for the NDS $\mathbf{\Sigma}_{p}$, it is necessary that its external output vector does not depend on its future external input vector, and each pair of admissible initial conditions and admissible excitation signals can generate one and only one response signal. From Lemma \ref{lemma:2}, this means that the descriptor system associated with the linear part of the NDS $\mathbf{\Sigma}_{p}$ must be regular. That is, the following assumption is necessary for investigating the formulated identification problem.

\begin{Assumption}\label{assum:1}
For each $\theta\in\mathbf{\Theta}$, the linear part of the NDS $\mathbf{\Sigma}_{p}$ is regular, meaning that the matrix valued polynomial (MVP) $s E-A(\theta)$ is invertible. Here, $s$ stands for the Laplace transform variable.
\end{Assumption}

Compared with an ordinary state-space model, a particular characteristic of a descriptor form model is that there may exist impulse modes in its time-domain responses, which is in general not appreciated in actual applications, noting that an impulse mode may significantly deteriorate system performances, and invalidate the linear approximation of the adopted descriptor form model that may even make the actual system unstable \cite{abg2020,Dai1989,Duan2010}. Based on these considerations, the following assumption is also adopted in this paper.

\begin{Assumption}\label{assum:9}
For each $\theta\in\mathbf{\Theta}$, the linear part of the NDS $\mathbf{\Sigma}_{p}$ is impulse free, meaning that the inverse of the MVP $s E-A(\theta)$ is proper.
\end{Assumption}

From Lemma \ref{lemma:2}, it is clear that when the linear part of the NDS $\mathbf{\Sigma}_{p}$ is regular and impulse free, then for each $k=2,3,\cdots$, the descriptor form model for $x_{k}(t)$ is also regular and impulse free. It can therefore be declared that when Assumptions \ref{assum:1} and \ref{assum:9} are satisfied simultaneously, both the state vector $x(t)$ and the output vector $y(t)$ of the NDS $\mathbf{\Sigma}_{p}$ are uniquely determined by its admissible initial states $x(0)$ and external input signal $u(t)$. Moreover, there do not exist any impulses in its state vector $x(t)$ and output vector $y(t)$.

In addition to these assumptions, the following assumptions are also introduced in this paper, which are helpful in avoiding awkward expressions. It is argued in the next section that these assumptions can be satisfied without significant difficulties, or can be easily removed in an actual identification problem.

\begin{Assumption}\label{assum:10}
For each $\theta\in\mathbf{\Theta}$, the generalized eigenvalues of the matrix pair $(E,\,A(\theta))$ in the linear part of the NDS $\mathbf{\Sigma}_{p}$ are different from each other. Moreover, all the eigenvalues of the state transition matrix $\Xi$ are also distinct from each other. In addition, each generalized eigenvalue of the matrix pair $(E,\,A(\theta))$ is distinct from any linear combination of the eigenvalues of the state transition matrix $\Xi$ of the assembly PSGS $\mathbf{\Sigma}_{s}$ with nonnegative integer coefficients.
\end{Assumption}

While Assumptions \ref{assum:2}, \ref{assum:1} and \ref{assum:9} are necessary for performing an identification experiment in open loop, Assumption \ref{assum:10} is adopted only for avoiding a complicated presentation that may hide the main ideas behind the suggested estimation procedures. As a matter of fact, all the results of this paper can be straightforwardly extended to the case when this assumption is not satisfied. But it is worthwhile to mention that when the matrix $\Xi$ is not similar to a diagonal matrix, derivatives of a TFM with respect to the Laplace variable exist in the response of the NDS $\mathbf{\Sigma}_{p}$, which may significantly complicate some  associated equations.

To simplify expressions, dependence of the system matrices of the NDS $ \mathbf{\Sigma}_{p} $, that is, $A(\theta)$, $B(\theta)$, $C(\theta)$, $D(\theta)$, $\Gamma_{x}(\theta)$ and $\Gamma_{u}(\theta)$, on its SIP vector $\theta$ is omitted in the rest of this section, as well as the generalized eigenvalues of the matrix pair $(E,\,A(\theta))$, and the TFM $H(s,\theta)$ for its linear part.

Denote the generalized eigenvalues of the matrix pair $(E,\,A)$ by $\lambda_{p,i}|_{i=1}^{m_{x}}$, while the eigenvalues of the matrix $\Xi$ by $\lambda_{s,i}|_{i=1}^{m_{\xi}}$. Moreover, for each $k \geq 1$ and $1 \leq i_{l} \leq m_{\xi}$ with $l=1,2,\cdots,k$, define a scalar $\lambda_{s}(i_{l}|_{l=1}^{k})$ as
\begin{displaymath}
\lambda_{s}(i_{l}|_{l=1}^{k}) = \sum_{l=1}^{k} \!\lambda_{s,i_{l}}
\end{displaymath}
Furthermore, for each $k \geq 2$, $1\leq l \leq k-1$, $1 \leq q \leq l$, and $1\leq i_{\star,h} \leq m_{\star}$ with $\star = x$ or $\xi$ and $1 \leq h \leq q$ or $l-q$ correspondingly, define a scalar $\lambda_{p,s}^{[l,q]}(i_{h}|_{h=1}^{l})$ as
\begin{displaymath}
\lambda_{p,s}^{[l,q]}(i_{h}|_{h=1}^{l}) = \sum_{h=2}^{q} \!\lambda_{p,i_{h}} \!+\! \sum_{h=q+1}^{l} \!\lambda_{s,i_{h}}
\end{displaymath}

Using these symbols, the following results are obtained on the basis of  Lemmas \ref{lemma:2} and \ref{lemma:3}, which give an explicit decomposition for the time-domain responses of the NDS $ \mathbf{\Sigma}_{p} $, that is expressed as the sum of those that are due to its initial conditions and its products with external stimulus, and those that are only due to external stimulus and their products. The former is usually called transient response of the NDS $ \mathbf{\Sigma}_{p} $, while the latter its steady-state response. The  proof is deferred to the appendix.

\newtheorem{Theorem}{Theorem}

\begin{Theorem}\label{theo:1}
Assume that the NDS $\mathbf{\Sigma}_{p}$ and the assembly PSGS $\mathbf{\Sigma}_{s}$ satisfy Assumptions \ref{assum:2}-\ref{assum:10} simultaneously. Then the output of the NDS $\mathbf{\Sigma}_{p}$ can be decomposed as
\begin{equation}
y(t) \!=\! C\! \sum_{k=1}^{\infty}\!\! x_{k}(t) \!+\! Du(t) \hspace{0.20cm}{\rm with} \hspace{0.2cm} x_{k}(t) \!=\! x_{k,t}(t) \!+\! x_{k,s}(t)
\label{eqn:theo:1-1}
\end{equation}
in which the vectors $x_{k,t}(t)$ and $x_{k,s}(t)$ have respectively the following expressions for each $k \geq 1$,
\begin{eqnarray}
& & \hspace*{-0.8cm}
x_{k,t}(t) \!=\!\! \sum_{i_{1}=1}^{m_{x}} \!e^{\lambda_{p,i_{1}}t} \!\!\left( \!\! \psi_{p}(i_{1}) \!+\!\!
\sum_{l=2}^{k} \!\sum_{q=1}^{l} \!\!\left\{ \!\!\sum_{i_{2}=1}^{m_{x}} \cdots \!\!\sum_{i_{q}=1}^{m_{x}}\sum_{i_{q+1}=1}^{m_{\xi}}
\cdots  \right.\right. \nonumber \\
& & \hspace*{0.2cm}
\left.\left.\sum_{i_{l}=1}^{m_{\xi}} \!\left[ \! e^{\lambda_{p,s}^{[l,q]}(i_{h}|_{h=1}^{l})t} \psi_{p,s}^{[l,q]}(i_{h}|_{h=1}^{l}) \right] \!\!\right\}\!\!\right) \label{eqn:theo:1-2} \\
& & \hspace*{-0.8cm}
x_{k,s}(t) \!=\!\! \sum_{i_{1}=1}^{m_{\xi}}\cdots\sum_{i_{k}=1}^{m_{\xi}}e^{\lambda_{s}(i_{h}|_{h=1}^{k})t}\psi_{s}(i_{h}|_{h=1}^{k})
\label{eqn:theo:1-3}
\end{eqnarray}
Here, for every associated admissible tuple of $i_{h}$s, $\psi_{p}(i_{1})$, $\psi_{p,s}^{[l,q]}(i_{h}|_{h=1}^{l})$ and $\psi_{s}(i_{h}|_{h=1}^{k})$ are some time independent vectors.
\end{Theorem}

In the above expressions, $x_{k,s}(t)$ only has dynamics that are linear combinations of the assembly PSGS $\mathbf{\Sigma}_{s}$ with nonnegative integer coefficients, while in $x_{k,t}(t)$, linear combinations are available for both the dynamics of the assembly PSGS $\mathbf{\Sigma}_{s}$ and the dynamics of the linear part of the NDS $\mathbf{\Sigma}_{p}$. This is different from that of a linear NDS, whose transient responses only contains the dynamics of the NDS itself, while its steady-state response only contains the dynamics of the PSGS  \cite{Zhou2025}. These linear combinations makes the associated interaction identification problem mathematically more difficult, noting that there are in general infinitely many such linear combinations. From the proof of Theorem \ref{theo:1}, it is clear that these linear combinations are resulted from the quadratic term $x(t)\otimes x(t)$ and the bilinear term $x(t)\otimes u(t)$ in the QBTI model, and represent high order harmonics in its time domain responses.

Note that for each $k=2,3,\cdots$, the transient state response $x_{k,t}(t)$ includes at least one mode in the dynamics of the linear part of the NDS $\mathbf{\Sigma}_{p}$. This means that when the linear part of the NDS $\mathbf{\Sigma}_{p}$ is stable and all the eigenvalues of the assembly PSGS $\mathbf{\Sigma}_{s}$ have a real part not greater than zero, then with the increment of the temporal variable $t$, $x_{k,t}(t)$ decreases exponentially to zero in magnitude for every $k \geq 2$. This leads possibilities of estimating the time independent vectors $\psi_{s}(i_{h}|_{h=1}^{k})$ in the steady-state response $x_{k,s}(t)$ of the NDS $\mathbf{\Sigma}_{p}$ from its input-output data. However, different from a linear NDS, existence of an eigenvalue with a positive real part in the assembly PSGS $\mathbf{\Sigma}_{s}$ may lead to the existence of a $k=2,3,\cdots$, such that the transient state response $x_{k,t}(t)$ increases exponentially in magnitude, that may prohibit estimations of a  tangential condition of the NDS $\mathbf{\Sigma}_{p}$. On the other hand, from the structure of the QBTI model, a recursive formula can be derived for these time independent vectors $\psi_{s}(i_{h}|_{h=1}^{k})$.

\newtheorem{Corollary}{Corollary}
\begin{Corollary} \label{coro:1}
Under the same assumptions of Theorem \ref{theo:1}, for each $k=2,3,\cdots$, and every tuple $i_{h}|_{h=1}^{k}$ with $i_{h}\in\{1,2,\cdots,m_{\xi}\}$, the time independent vectors $\psi_{s}(i_{h}|_{h=1}^{k})$ of Equation (\ref{eqn:theo:1-3}) can be recursively expressed as
\begin{eqnarray}
& & \hspace*{-1.0cm} \psi_{s}(i_{h}|_{h=1}^{k}) \!=\!\!\! \left[ \!\!\left( \!\sum_{h=1}^{k} \!\!\lambda_{s,i_{h}} \!\!\!\right) \!\!E \!-\! A \!\right]^{\!-\!1} \!\!\!\! \left\{ \!\Gamma_{x}  \! \!\left[ \sum_{l=1}^{k-1}\psi_{s}(i_{h}\!|_{h=1}^{l}) \otimes  \right.\right. \nonumber\\
& & \hspace*{0.8cm}  \left.\left. \psi_{s}(i_{h}\!|_{h=l+1}^{k}) \!\right] \! \!+ \!  \Gamma_{u} \! \!\left[\psi_{s}(i_{h}\!|_{h=1}^{k-1})  \!\otimes \! \psi_{u}(i_{k}) \!\right] \!\right\}
\label{eqn:coro:1-1}
\end{eqnarray}
with $\psi_{s}(i) \!=\! (\lambda_{s,i}E \!-\!A)^{\!-\!1} \!B \psi_{u}(i)$ for each $i=1,2,\cdots,m_{\xi}$.
\end{Corollary}

A proof of this corollary is given in the appendix.

For each $k=1,2,\cdots$, let $s_{i}$ with $i=1,2,\cdots,k$, denote the Laplace variable of the $k$-th dimensional Laplace transform. Define a TFM $G(s_{1})$ as
\begin{equation}
G(s_{1}) = (s_{1}E -A)^{-1} B
\end{equation}
and a multiple dimensional TFM $G(s_{i}|_{i=1}^{k})$, which is sometimes also called as a generalized TFM \cite{abg2020,kga2022},  with $k\geq 2$ as
\begin{eqnarray}
& & \hspace*{-1.3cm} G(s_{i}|_{i=1}^{k}) \!=\!\!\! \left[ \!\!\left( \!\sum_{i=1}^{k} \!\! s_{i} \!\!\!\right) \!\!E \!-\! A \!\right]^{\!-\!1} \!\!\!\! \left\{ \!\Gamma_{x}  \! \!\left[ \sum_{l=1}^{k-1}G(s_{i}\!|_{i=1}^{l}) \otimes  \right.\right. \nonumber\\
& & \hspace*{0.8cm}  \left.\left. G(s_{i}\!|_{i=l+1}^{k}) \!\right] \! \!+ \!  \Gamma_{u} \! \!\left[G(s_{i}\!|_{i=1}^{k-1})  \!\otimes \! I_{m_{u}} \!\right] \!\right\}
\label{eqn:coro:1-2}
\end{eqnarray}
Then from Equation (\ref{app:1-17}), it can be straightforwardly shown that for each tuple $i_{h}|_{h=1}^{k}$ with $i_{h}=1,2,\cdots,m_{\xi}$, and $k=1,2,\cdots$, we have that
\begin{equation}
\hspace*{-0.0cm} \psi_{s}(i_{h}|_{h=1}^{k}) \!=\!
G(\lambda_{s,i_{h}}|_{h=1}^{k}) \psi_{u}(i_{h}|_{h=1}^{k})
\label{eqn:coro:1-3}
\end{equation}
in which $\psi_{u}(i_{h}|_{h=1}^{k}) = \psi_{u}(i_{1}) \!\otimes\! \psi_{u}(i_{2}) \!\otimes\!\cdots \!\otimes\! \psi_{u}(i_{k})$.

These results can be extended to the case in which the PSGS itself is also a QBTI system.

\section{Nonparametric and Parametric Estimation with a Multi-sine Probing Signal}\label{sec:par-est}

In the previous section, an explicit formula is given for the response of the NDS $\mathbf{\Sigma}_{p}$ under the stimulation of the output of an LTI system. Different from that of an LTI NDS, in this response, not only the modes of the NDS $\mathbf{\Sigma}_{p}$ and the assembly PSGS $\mathbf{\Sigma}_{s}$, but also their combinations with some nonnegative integer coefficients are also included. This makes the associated NDS interaction estimation mathematically more involved.

To deal with the interaction identification problem,
define a TFM $H(s_{1},\theta)$ and a multiple dimensional/generalized TFM $H(s_{i}|_{i=1}^{k},\theta)$ with $k=2,3,\cdots$, respectively as follows,
\begin{eqnarray}
& & \hspace*{-1.0cm}
H(s_{1},\theta) = C(\theta)\left[ s_{1}E-A(\theta)\right]^{-1}B(\theta) + D(\theta)
\label{eqn:tfm-1}\\
& & \hspace*{-1.0cm}
H(s_{i}|_{i=1}^{k},\theta) \!=\!C(\theta) G(s_{i}|_{i=1}^{k},\theta)
\label{eqn:tfm-2}
\end{eqnarray}
in which $G(s_{i}|_{i=1}^{k},\theta)$ is defined by Equation (\ref{eqn:coro:1-2}), explicitly expressing its dependence on the SIP vector $\theta$.

Denote $C(\theta)\sum_{k=1}^{\infty}\!\! x_{k,t}(t)$ and $C(\theta)\sum_{k=1}^{\infty}\!\! x_{k,s}(t) \!+\! D(\theta)u(t)$ respectively by $y_{t}(t,\theta)$ and $y_{s}(t,\theta)$, standing respectively for the transient response and the steady-state response of the NDS $\mathbf{\Sigma}_{p}$. Then it can be directly claimed from Theorem \ref{theo:1} that the output vector $y(t,\theta)$ of the NDS $\mathbf{\Sigma}_{p}$ can be expressed as
\begin{equation}
y(t,\theta) = y_{t}(t,\theta) + y_{s}(t,\theta)
\end{equation}
In addition, from Corollary \ref{coro:1}, as well as the definitions of the generalized TFM $H(s_{i}|_{i=1}^{k},\theta)$, it is obvious that the steady-state response $y_{s}(t,\theta)$ has the following representations,
\begin{equation}
y_{s}(t,\theta) \!=\!\! \sum_{k=1}^{\infty} \!\sum_{i_{1}=1}^{m_{\xi}} \!\!\cdots\!\! \sum_{i_{k}=1}^{m_{\xi}} \!\! e^{\lambda_{s}(i_{h}|_{h=1}^{k})t}\phi_{u}(i_{h}|_{h=1}^{k},\theta)
\label{eqn:resdec-1}
\end{equation}
in which
\begin{equation}
\phi_{u}(i_{h}|_{h=1}^{k},\theta) = H(\lambda_{s,i_{h}}|_{h=1}^{k},\theta)
\psi_{u}(i_{h}|_{h=1}^{k})
\label{eqn:resdec-2}
\end{equation}

Note that for each $k=1,2,\cdots$, both $H(\lambda_{s,i_{h}}|_{h=1}^{k},\theta)$ and $\psi_{u}(i_{h}|_{h=1}^{k})$ does not depend on the temporal variable $t$, and are respectively a constant matrix and a constant vector when the index variables $i_{h}|_{h=1}^{k}$ are given. On the other hand, the above equation makes it clear that the steady-state response of the NDS $\mathbf{\Sigma}_{p}$ depends linearly on $H(\lambda_{s,i_{h}}|_{h=1}^{k},\theta)$. Moreover, $\phi_{u}(i_{h}|_{h=1}^{k},\theta)$ is called as a (right) tangential condition in operator theories and system analysis and synthesis, etc. \cite{bgr1990}, and plays important roles in system identification and model reduction \cite{abg2020,bgd2020,msa2024}. These relations are very similar to those of a linear NDS revealed in \cite{Zhou2025}, and make it possible to divide the NDS interaction identification into two stages, that is, a nonparametric estimation stage followed by a parametric estimation stage.

To solve this problem, the following algebraic results are required, which can be straightforwardly proved through some simple algebraic manipulations using the Euler formulas $cos(\phi)=(e^{\mathbf{i}\phi} + e^{-\mathbf{i}\phi})/2$ and $sin(\phi)=(e^{\mathbf{i}\phi} - e^{-\mathbf{i}\phi})/(2\mathbf{i})$, in which $\phi$ is an arbitrary real number. The proof is therefore omitted.

\begin{Lemma}
Let $\alpha$ and $\beta$ be some real numbers, while $n$ be a positive integer. Define $S(n,\alpha, \beta)$ as
\begin{displaymath}
S(n,\alpha, \beta) = \sum_{k=0}^{n}e^{k(\alpha + \mathbf{i} \beta)}
\end{displaymath}
Then for an arbitrary positive integer $n$, its real part $S^{[r]}(n,\alpha, \beta)$ and imaginary part $S^{[i]}(n,\alpha, \beta)$ can be respectively given by
\begin{eqnarray}
& & \hspace*{-1.1cm} S^{[r]}(n,\alpha, \beta) \nonumber \\
& & \hspace*{-1.5cm} =\!\! \frac{e^{(n \!+\! 2)\alpha} \!cos(n\beta) \!- e^{(n \!+\! 1)\alpha} \!cos[(n \!+\! 1)\beta] \!-\! e^{\alpha} \!cos(\beta) \!+\! 1}
{(e^{\alpha }-1)^{2}+4sin^{2}(\beta/2)}
\label{eqn:lemma:4-1} \\
& & \hspace*{-1.1cm} S^{[i]}(n,\alpha, \beta) \nonumber \\
& & \hspace*{-1.5cm} =\!\! \frac{e^{(n \!+\! 2)\alpha} \!sin(n\beta) \!- e^{(n \!+\! 1)\alpha} \!sin[(n \!+\! 1)\beta] \!+\! e^{\alpha} \!sin(\beta)}
{(e^{\alpha}-1)^{2}+4sin^{2}(\beta/2)}
\label{eqn:lemma:4-2}
\end{eqnarray}
\label{lemma:4}
\end{Lemma}

Note that when $\alpha=0$ and $\beta=2l\pi$ with $l=0,\pm 1, \pm 2,\cdots$, $(e^{\alpha} \!-\! 1)^{2} \!+\! 4sin^{2}(\beta/2) \!=\! 0$, meaning that the right hand sides of Equations (\ref{eqn:lemma:4-1}) and (\ref{eqn:lemma:4-2}) may not be well defined. Recall that a sinusoidal function is a periodic function. Direct algebraic manipulations show that under such a situation, both $S^{[r]}(n,\alpha, \beta)$ and $S^{[i]}(n,\alpha, \beta)$ can be defined as its limit with $\alpha=0$ and $\beta$ approaching zero.

From this definition and the above Lemma, it is clear that for an arbitrary integer $l$, the following relations are valid.
\begin{eqnarray}
& & \hspace*{-1.0cm}
 \lim_{n\rightarrow\infty}\frac{S^{[r]}(n,\alpha, \beta)}{n+1} = \left\{\begin{array}{ll}
0 & \alpha < 0  \\
0 & \alpha = 0,\;\; \beta \neq 2l\pi  \\
1 & \alpha = 0,\;\; \beta = 2l\pi  \end{array} \right. \\
& & \hspace*{-1.0cm}
\lim_{n\rightarrow\infty}\frac{S^{[i]}(n,\alpha, \beta)}{n+1} = 0,\;\; \alpha\leq 0
\end{eqnarray}

To develop an estimation algorithm, the following assumption is introduced.

\begin{Assumption}\label{assum:11}
For each $\theta\in\mathbf{\Theta}$, the linear part of the NDS $\mathbf{\Sigma}_{p}$ is stable. Moreover, the assembly PSGS $\mathbf{\Sigma}_{s}$ has all its eigenvalues on the imaginary axis that are distinct from each other, and each of its first $m_{\xi,+}$ ones has a nonnegative imaginary part, while each of the remaining has a negative imaginary part. In addition, there does not exist any tuple of nonnegative integers $k_{i}|_{i=1}^{m_{\xi,+}}$ that are not simultaneously equal to zero, such that $\sum_{i=1}^{m_{\xi,+}}k_{i}\lambda_{s,i} = 2l\pi$ with $l$ being an arbitrary nonnegative integer.
\end{Assumption}

It is worthwhile to mention that while the stability of the linear part of the NDS $\mathbf{\Sigma}_{p}$ is extensively regarded as necessary to perform an open-loop identification experiment, the assumptions on the assembly PSGS $\mathbf{\Sigma}_{s}$ are general not. These assumptions are due to the existence of mixed modes in the steady-state response of the NDS $\mathbf{\Sigma}_{p}$, that bring mathematical difficulties to the  nonparametric estimation stage.

When the linear part of the NDS $\mathbf{\Sigma}_{p}$ is stable, we have that $\lambda_{p,i}^{[r]} < 0$ for each $i=1,2,\cdots,m_{x}$. Recall that for every positive integers $k$ and $l$ with $l \leq k$, as well as every feasible tuple of $i_{h}|_{h=1}^{k}$, $\psi_{p}(i_{1})$, $\psi_{p,s}^{[l,q]}(i_{h}|_{h=1}^{l})$ and $\psi_{s}(i_{h}|_{h=1}^{k})$ are time independent. It is clear from Equation (\ref{eqn:theo:1-2}) that when Assumption \ref{assum:11} is satisfied, all the transient responses $x_{k,t}(t)$ with $k=1,2,\cdots$, decay exponentially to zero in magnitude.

For each tuple $ i_{h}|_{h=1}^{l}$ with $i_{h} \in \{1,2,\cdots, m_{\xi,+}\}$, define a set ${\cal S}^{[r]}\{i_{h}|_{h=1}^{l}\}$ and a set ${\cal S}^{[i]}\{i_{h}|_{h=1}^{l}\}$ respectively as
\begin{eqnarray*}
& & \hspace*{-0.6cm} {\cal S}^{[r]}\{i_{h}|_{h=1}^{l}\} = \left\{ i_{g}|_{g=1}^{q} \left| \begin{array}{l}
for\;\; each\;\; k\in{\cal N},\; there\\ exists\;\; a \;\;
p \in{\cal N}_{+},\;\; such\\ that \;\; i_{g}|_{g=1}^{q}=\frac{2p\pi}{kT} \pm i_{h}|_{h=1}^{l} \end{array}\right. \right\}
 \\
& & \hspace*{-0.6cm} {\cal S}^{[i]}\{i_{h}|_{h=1}^{l}\} \!=\! \left\{\! i_{g}|_{g=1}^{q} \left| \begin{array}{l}
for\;\; each\;\; k\in{\cal N},\; there\\ exists\;\; a \;\;
p \in{\cal N}_{+},\;\; such\\ that \;\; i_{g}|_{g=1}^{q} \!=\!\frac{p\pi}{kT} \!+\! (-1)^{p} i_{h}|_{h=1}^{l} \end{array}\right. \!\!\!\!\right\}
\end{eqnarray*}
in which ${\cal N}$ and ${\cal N}_{+}$ stands respectively for the set consisting of nonnegative and positive integers.
Then in addition to the above observations, we also have the following conclusions which is greatly helpful in nonparametric estimation for the NDS $\mathbf{\Sigma}_{p}$. Their proof is included in the appendix.

\begin{Theorem}
Assume that the NDS $\mathbf{\Sigma}_{p}$ and the assembly PSGS $\mathbf{\Sigma}_{s}$ satisfy simultaneously Assumptions \ref{assum:2}-\ref{assum:11}. Then for each sampling period $T$, as well as for each $l \geq 1$ and $ i_{h} \in \{1,2,\cdots, m_{\xi,+} \}$ with $h=1,2,\cdots,l$, we have the following equalities,
\begin{eqnarray}
& & \hspace*{-0.2cm} \lim_{n \rightarrow \infty} \!\!\frac{1}{n \!+\! 1} \!\!\sum_{k=0}^{n} \!\!cos \!\left(\!k\lambda_{s}^{[i]}\!(i_{h}|_{h=1}^{l})T \!\right) \! y(kT, \theta) \nonumber \\
& & \hspace*{-0.8cm}\!=\! \sum_{i_{g}|_{g=1}^{q}\in {\cal S}^{[r]}\{i_{h}|_{h=1}^{l}\} } \hspace{-0.4cm}   \phi_{u}^{[r]}\!(i_{g}|_{g=1}^{q}\!,\theta) \\
& & \hspace*{-0.2cm} \lim_{n\rightarrow \infty} \!\!\frac{-1}{n \!+\! 1} \!\!\sum_{k=0}^{n} \!\!sin \!\left( \!k \lambda_{s}^{[i]}\!(i_{h}|_{h=1}^{l})T \!\right)\! y(kT, \theta) \nonumber \\
& & \hspace*{-0.8cm}\!=\! \sum_{i_{g}|_{g=1}^{q}\in {\cal S}^{[i]}\{i_{h}|_{h=1}^{l}\} } \hspace{-0.4cm} \phi_{u}^{[i]}\!(i_{h}|_{h=1}^{q}\!,\theta)
\end{eqnarray}
\label{theo:2}
\end{Theorem}

From Equation (\ref{eqn:resdec-2}), it is clear that ${\phi}_{u}\!(i_{h}|_{h=1}^{l})$ is actually the value of the TFM $H(s_{h}|_{h=1}^{k},\theta)$ at $s_{h} = \lambda_{s,i_{h}}$, $h=1,2,\cdots,k$, along the direction $\psi_{u}(i_{1}) \otimes \psi_{u}(i_{2}) \otimes \cdots \otimes \psi_{u}(i_{k})$. This value is usually called a tangential interpolation of the TFM $H(s_{h}|_{h=1}^{k},\theta)$, that is widely used in model reduction, system identification and functional analysis, etc. \cite{abg2020,bgd2020,Gosea2022,msa2024}. Different from the results of \cite{Zhou2025} in which each subsystem of the NDS $\mathbf{\Sigma}_{p}$ is linear, it appears from Theorem \ref{theo:2} that estimation for a tangential interpolation condition ${\phi}_{u}\!(i_{h}|_{h=1}^{l})$ of a QBTI system is in general quite complicated and challenging, noting that there are usually infinitely many elements in the sets ${\cal S}^{[r]}\{i_{h}|_{h=1}^{l}\} $ and ${\cal S}^{[i]}\{i_{h}|_{h=1}^{l}\} $. This is due to the existence of multiplications among system states, as well as those between a system state and an external input.

However, if $\bar{\sigma}\{[( \sum_{i=1}^{k} \!\! s_{i}) E - A ]^{\!-\!1}\}$ is less than $1$, then from Equations (\ref{eqn:coro:1-2}), (\ref{eqn:tfm-2}) and (\ref{eqn:resdec-2}), it can be directly declared that $\left|\left|\phi_{u}\!(i_{h}|_{h=1}^{q}\!,\theta)\right|\right|$ decreases at least exponentially with the increment of $q$, meaning that when $q \in {\cal N}_{+} $ is sufficiently large, $\left|\left|\phi_{u}\!(i_{h}|_{h=1}^{q}\!,\theta)\right|\right|$ is very small. This makes it possible to estimate approximately the aforementioned tangential interpolation conditions, that are associated with some fundamental frequencies in the steady-state response of the QBTI system, as well as those of low order harmonic frequencies.

More specifically, assume that $\bar{\sigma}\{[( \sum_{i=1}^{k} \!\! s_{i}) E - A ]^{\!-\!1}\} \ll 1$. Then based on Theorem \ref{theo:2}, an approximate estimate for $\phi_{u}\!(i_{h}|_{h=1}^{q},\theta)$, denote it by $\widehat{\phi}_{u}\!(i_{h},\theta)$, can be directly obtained for any prescribed fundamental frequency $\lambda_{s,i_{h}}$ with $ i_{h} \in \{1,2,\cdots, m_{\xi,+} \}$. Particularly, let $y_{m}(kT)$ denote the measured value of the external output vector $y(t)$ of the NDS $\mathbf{\Sigma}_{p}$ at the sampling instant $kT$, $k=0,1,\cdots,N_{d}$. Then
\begin{equation}
 \hspace*{-1.4cm} \widehat{\phi}_{u}\!(i_{h}|_{h=1}^{q},\theta) \approx
\frac{1}{n \!+\! 1} \!\sum_{k=0}^{N_{d}} \!e^{\mathbf{i}k\lambda_{s}^{[i]}\!(i_{h}|_{h=1}^{q})T} \! y_{m}(kT)
\label{eqn:non-par-est-1}
\end{equation}

It is also worthwhile to mention that different from the results of \cite{Zhou2025}, in the above estimation procedure, a tangential condition can be estimated for the generalized TFMs of the NDS $\mathbf{\Sigma}_{p}$ only at the imaginary axis. In addition, output sampling is required to be periodic. This is mainly due to the combinations of the dynamics of the linear part of the NDS $\mathbf{\Sigma}_{p}$ and those of the assembly PSGS $\mathbf{\Sigma}_{s}$, that is once again caused by the multiplications among system states and those between a system state and an external input.

To recover the value of the SIP vector $\theta$ from an estimate of the tangential interpolation of the TFM $H(s_{h}|_{h=1}^{k},\theta)$, the following results are derived, while their proof is given in the appendix.

\begin{Theorem}
For each $k=1,2,\cdots$, the generalized TFM $H(s_{i}|_{i=1}^{k},\theta)$ depends on the SIP vector $\theta$ through an LFT.
\label{theo:3}
\end{Theorem}

The above theorem makes it clear that for every $k \geq 1$ and any tuple $i_{h}|_{h=1}^{k}$, the MVF $H(\lambda_{s,i_{h}}|_{h=1}^{k},\theta)$, and therefore the vector $\phi_{u}\!(i_{h}|_{h=1}^{k}\!,\theta)$, depends through an LFT on the NDS SIP vector $\theta$, recalling that the vector $\psi_{u}(i_{h}|_{h=1}^{k})$ is completely determined by the assembly PSGS $\mathbf{\Sigma}_{s}$ for a fixed tuple $i_{h}|_{h=1}^{k}$.

On the basis of this relation between $\phi_{u}\!(i_{h}|_{h=1}^{k}\!,\theta)$ and $\theta$, as well as the observation that any addition of LFTs can still be expressed as an LFT \cite{zdg1996}, through similar derivations as those of \cite{Zhou2025}, an estimate $\widehat{\theta}$ can be obtained for the SIP vector $\theta$ from an approximate estimate of some  $\phi_{u}\!(i_{h}|_{h=1}^{k}\!,\theta)$s, that is, $\widehat{\phi}_{u}\!(i_{h}|_{h=1}^{k}\!,\theta)$s, or straightforwardly from an estimate of
\begin{displaymath}
\sum_{i_{g}|_{g=1}^{q}\in {\cal S}^{[r]}\{i_{h}|_{h=1}^{k}\} } \hspace{-1.0cm}   \phi_{u}^{[r]}\!(i_{g}|_{g=1}^{q}\!,\theta)
\hspace{0.25cm} {\rm and/or} \hspace{0.25cm}
\sum_{i_{g}|_{g=1}^{q}\in {\cal S}^{[i]}\{i_{h}|_{h=1}^{k}\} } \hspace{-1.0cm} \phi_{u}^{[i]}\!(i_{h}|_{h=1}^{q}\!,\theta)
\end{displaymath}

In summary, the estimation for the SIP vector $\theta$ consists of the following two steps.

\begin{itemize}
\item Nonparametric Estimation. Select a set of appropriate integers $i_{h}(l)|_{h=1}^{q(l)}$ with $l=1,2,\cdots,N_{e}$ and $i_{h}(l) \in \{1,2,\cdots,m_{\xi,+}\}$. Estimate the tangential interpolation condition  ${\phi}_{u}\!(i_{h}(l)|_{h=1}^{q(l)},\theta)$ using Equation (\ref{eqn:non-par-est-1}).
\item Parametric Estimation. Estimate the SIP vector $\theta$ from $\widehat{\phi}_{u}\!(i_{h}(l)|_{h=1}^{q(l)},\theta)$ with $l=1,2,\cdots,N_{e}$, using the least squares data fitting techniques of \cite{Zhou2025}, on the basis of the LFT expression of Theorem \ref{theo:3}.
\end{itemize}

By the same token of \cite{Zhou2025}, several statistical properties, such as convergence, etc., can be established under some weak assumptions on measurement errors, etc., respectively for the nonparametric estimate $\widehat{\phi}_{u}\!(i_{h}|_{h=1}^{l}\!,\theta)$ and the parametric estimate $\widehat{\theta}$. For example, if we denote the composite influences of process disturbances, measurement errors, etc., on the external output vector $y(t,i)$ of the NDS subsystem $\mathbf{\Sigma}_{p,i}$, in which $i=1,2,\cdots,N_{p}$, by a time series $n(t,i)$. Then these properties can be guaranteed under the condition that $n(t,i)$ is uncorrelated at each sampling time instant $t_{k}$ with $k=1,2,\cdots,N_{d}$, $n(t_{ki},i)$ and $n(t_{kj},j)$ are uncorrelated whenever $i\neq j$, the expectation of $n(t_{k},i)$ is equal to zero, while its covariance matrix is not greater than a constant positive definite matrix that has a finite maximum singular value.

It is worthwhile to emphasize that in order to guarantee that information is efficiently utilized in the aforementioned estimations, that is contained in the measurements of the sampled NDS external outputs, the associated  $\lambda_{s,i_{h}}|_{h=1}^{l}$s must be appropriately selected. Further efforts are required to settle this selection issue.

\section{A Numerical Example}\label{sec:num-exam}

To illustrate characteristics of the suggested estimation algorithm, this section considers parameter identification for a simple circuit consisted from 2 capacitors and 2 diodes, which is also adopted in \cite{kga2022} to demonstrate how to convert the model of a nonlinear dynamic system into a QBTI model. These capacitors and diodes are divided into two blocks that are connected in series, and the input of the circuit is a current, while the output consists of voltage drops of each capacitor. Figure 1 gives the structure of this circuit, as well as the associated  relations between inputs and outputs of its diodes.

\renewcommand{\thefigure}{\arabic{figure}}
\setcounter{figure}{0}
\vspace{-0.0cm}
\begin{figure}[t]
\vspace{-0.4cm}
\begin{center}
\includegraphics[width=2.0in]{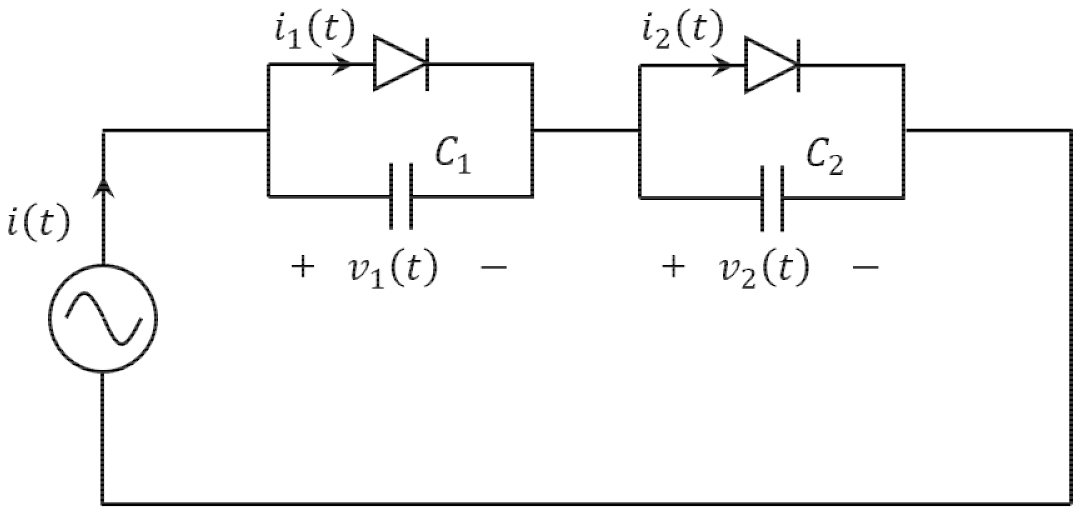}
\hspace{0.35cm}
\includegraphics[width=1.0in]{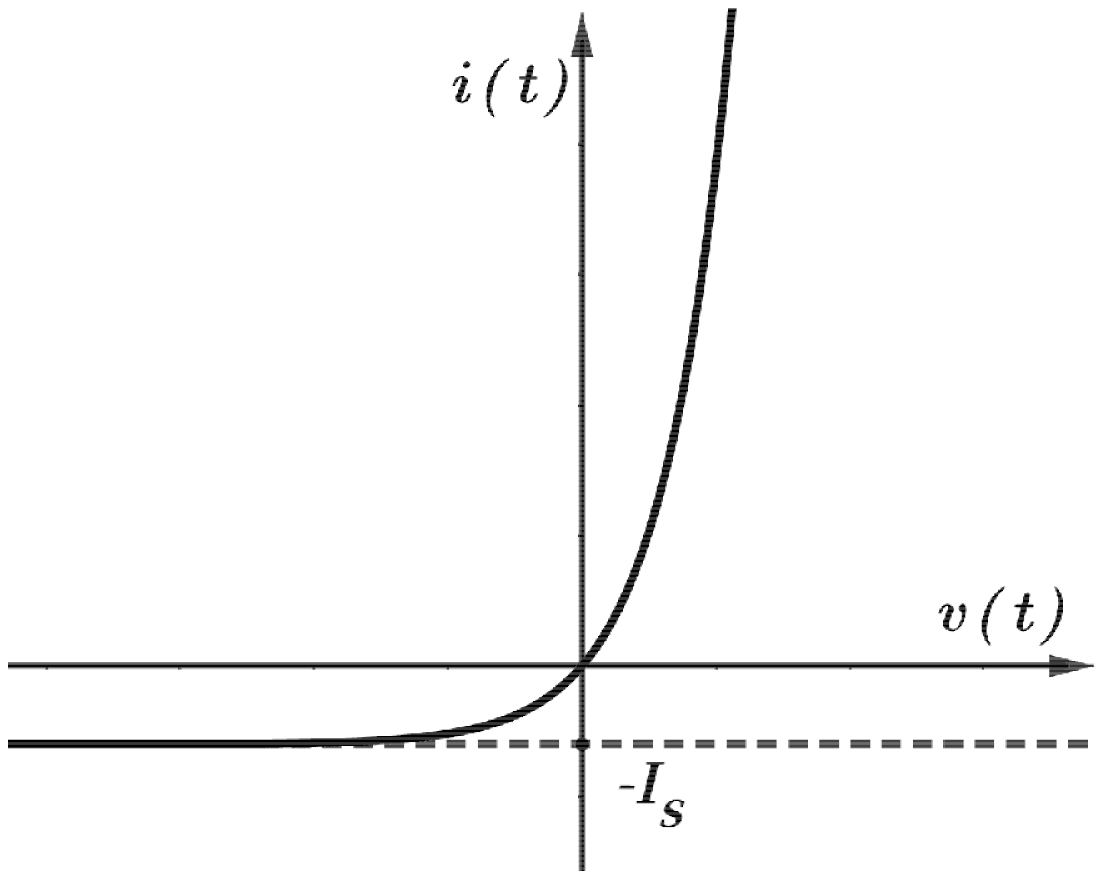}

\hspace*{1.5cm} (a) the circuit \hspace{1.5cm} (b) diode characteristics

\vspace{-0.4cm}\hspace*{5cm} \caption{Structure of the Circuit and Input-output Properties of a Diode.}

\label{fig:1}
\vspace{-0.4cm}
\end{center}
\end{figure}

For each $i=1,2$, let $v_{i}(t)$ represent the voltage drop of the $i$-th capacitor. Then according to working principles of the circuit, the following dynamic models can be established \cite{kga2022},
\begin{equation}
C_{i}\frac{dv_{i}(t)}{dt} = i(t) - I_{s,i}\left( e^{v_{i}(t)/V_{th,i}} - 1 \right),\hspace{0.25cm} i=1,2
\label{eqn:num-ex-1}
\end{equation}
in which $C_{i}$, $I_{s,i}$ and $V_{th,i}$ stand respectively for the value of the $i$-th capacitor, the saturation current and the temperature equivalent voltage of the $i$-th diode.

To be consistent with the symbols in the adopted QBTI model, denote $i(t)$ and $v_{i}(t)$ with $i \in \{1,\; 2\}$ respectively by $u(t)$ and $y_{i}(t)$. Moreover, introduce state variables $x_{i}(t)|_{i=1}^{6}$ for this circuit respectively as,
\begin{eqnarray*}
& & \hspace*{-1.0cm} x_{1}(t)=v_{1}(t), \hspace{0.3cm}
x_{2}(t)=i_{c1}(t), \hspace{0.3cm}
x_{3}(t) = e^{v_{1}(t)/V_{th,1}} - 1  \\
& & \hspace*{-1.0cm} x_{3}(t)=v_{2}(t), \hspace{0.3cm} x_{4}(t)=i_{c2}(t), \hspace{0.3cm} x_{4}(t) = e^{v_{2}(t)/V_{th,2}} - 1
\end{eqnarray*}
Then the following QBTI system model can be obtained from Equation (\ref{eqn:num-ex-1}), which is equivalent to the original system in the sense that the input-output relations remain unchanged \cite{kga2022}.
\begin{eqnarray*}
& &\hspace*{-1.0cm} \frac{dx_{1}(t)}{dt}=  \frac{1}{C_{1}} x_{2}(t)    \\
& &\hspace*{-1.0cm} 0 = x_{2}(t) + I_{s,1} x_{3}(t) - u(t)   \\
& & \hspace*{-1.0cm} \frac{dx_{3}(t)}{dt}= \frac{1}{C_{1}V_{th,1}} x_{2}(t) + \frac{1}{C_{1}V_{th,1}} x_{2}(t)x_{3}(t)  \\
& &\hspace*{-1.0cm} \frac{dx_{4}(t)}{dt}=  \frac{1}{C_{2}} x_{5}(t)    \\
& &\hspace*{-1.0cm} 0 = x_{5}(t) + I_{s,2} x_{6}(t) - u(t)   \\
& & \hspace*{-1.0cm} \frac{dx_{6}(t)}{dt}= \frac{1}{C_{2}V_{th,2}} x_{5}(t) + \frac{1}{C_{2}V_{th,2}} x_{5}(t)x_{6}(t)  \\
& & \hspace*{-1.0cm} y(t)=\left[\begin{array}{cccccc} 1 & 0 & 0 & 0 & 0 & 0\\ 0 & 0 & 0 & 1 & 0 & 0\end{array}\right] x(t)
\end{eqnarray*}

Assume that the values of $C_{i}|_{i=1}^{2}$ and $I_{s,i}|_{i=1}^{2}$ are known for this circuit. The objectives of this numerical example is to estimate the values of the parameters $V_{th,i}|_{i=1}^{2}$ from experiment data, that is, $\theta = [V_{th,1}\;\; V_{th,2}]^{T}$.

It is worthwhile to mention that a diode is widely used in modelling other electronic elements and devices, such as a photovoltaic cell, and its parameters are believed extensively hard to be estimated accurately due to the involved nonlinearities \cite{prddv2025}.

On the basis of the above equations, straightforward algebraic manipulations show that they can be equivalently expressed by the QBTI model of
Equations (\ref{plant-1}) and (\ref{plant-2}), in which all system matrices depend through an LFT on the parameter vector $\theta$. In addition, its (generalized) TFMs can be expressed as
\begin{eqnarray}
& &\hspace*{-1.0cm} H(s_{1},\theta) = \left[\frac{V_{th,1}}{C_{1}V_{th,1}s_{1} + I_{s,1} } \;\;\; \frac{V_{th,2}}{C_{2}V_{th,2}s_{1} + I_{s,2} } \right]^{T}   \label{eqn:num-sim-1}   \\
& &\hspace*{-1.0cm} H(s_{1},s_{2},\theta) =-\frac{s_{1}}{s_{1}+s_{2}} \left[ \begin{array}{c} H(s_{1},s_{2},\theta, 1)
 \\ H(s_{1},s_{2},\theta, 2)
\end{array}\right]    \label{eqn:num-sim-2}
\end{eqnarray}
in which for each $i=1,2$,
\begin{eqnarray*}
H(s_{1},s_{2},\theta,i) \!\!&=&\!\!
\frac{I_{s,i}V_{th,i}}{C_{i}V_{th,i}(s_{1}+ s_{2})+ I_{s,i}}\times \\
& & \frac{1}{C_{i}V_{th,i}s_{1} + I_{s,i}} \times
\frac{1}{C_{i}V_{th,i}s_{2} + I_{s,i}}
\end{eqnarray*}
Once again, each of these generalized TFMs is an LFT of the parameter vector $\theta$.

From these expressions, it is clear that the linear part of this circuit, which is represented by the TFM $H(s_{1},\theta)$, is stable, noting that all the involved physical parameters take a positive value. Moreover, it can be straightforwardly shown that the parameter vector $\theta$ is identifiable with the value of the TFM $H(s_{1},\theta)$ at {\it only} a single frequency point. In addition, when a tangential interpolation condition of the generalized TFM $H(s_{1},s_{2},\theta)$ is to be used in estimating the value of the parameter vector $\theta$, a tangential interpolation condition for the TFM $H(s_{1}+s_{2},\theta)$ is introduced for this numerical example, in order to satisfy the conditions required in \cite{Zhou2025} for getting a parametric estimate through a least squares data fitting.

On the basis of these observations, a numerical identification experiment is designed, in which a probing signal $u(t)=5  sin(\omega_{0}t)$ is added to the circuit. More precisely, system matrices of the PSGS $\mathbf{\Sigma}_{s}$, as well as its initial conditions, are selected as follows
\begin{displaymath}
\Xi=\left[\begin{array}{cc} 0 & \omega_{0} \\ -\omega_{0} & 0 \end{array}\right], \hspace{0.25cm}
\Pi=\left[\begin{array}{cc} 2.5 & -2.5 \end{array}\right], \hspace{0.25cm}
\xi(0)=\left[\begin{array}{c} 1 \\ 1 \end{array}\right]
\end{displaymath}

It is worthwhile to point out that in this circuit, $V_{th,1}$ and $V_{th,2}$ can be {\it independently} estimated from $y_{1}(t)$ and $y_{2}(t)$ respectively. This can be understood without significant difficulties from the aforementioned model of the circuit. This circuit is chosen mainly for illustrating influences of the decaying factor $\bar{\sigma}\{[( \sum_{i=1}^{k} \!\! s_{i}) E - A ]^{\!-\!1}\}$, which is actually $[C_{i}V_{th,i}( \sum_{j=1}^{k} \!\! s_{j}) + I_{s,i}]^{-1}$ with $i=1,2$, in this numerical example, on accuracies of the associated parametric and nonparametric estimations.

\begin{figure}[t]
\vspace{-0.0cm}
\begin{center}

\includegraphics[width=1.5in]{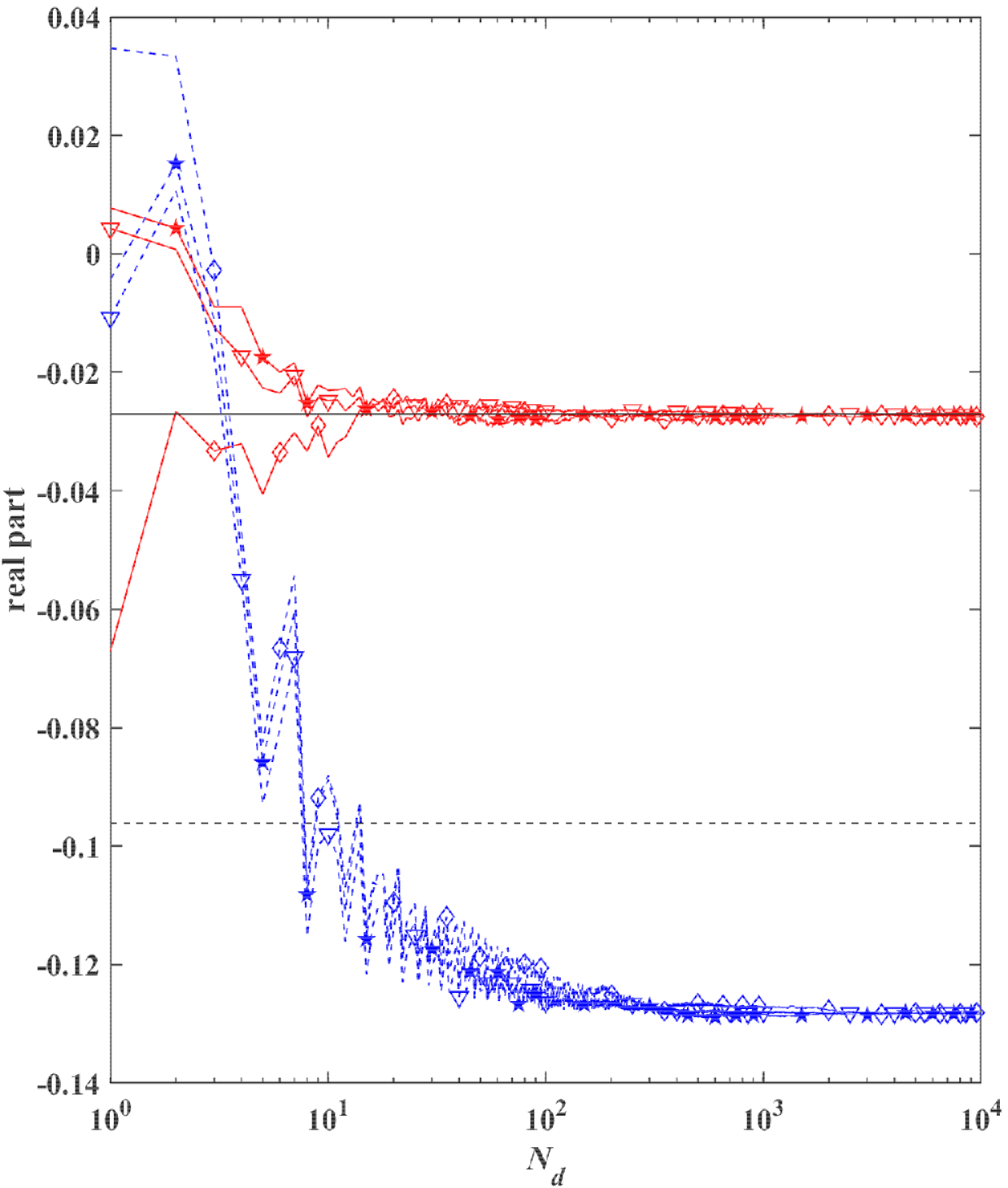}
\hspace{0.35cm}
\includegraphics[width=1.5in]{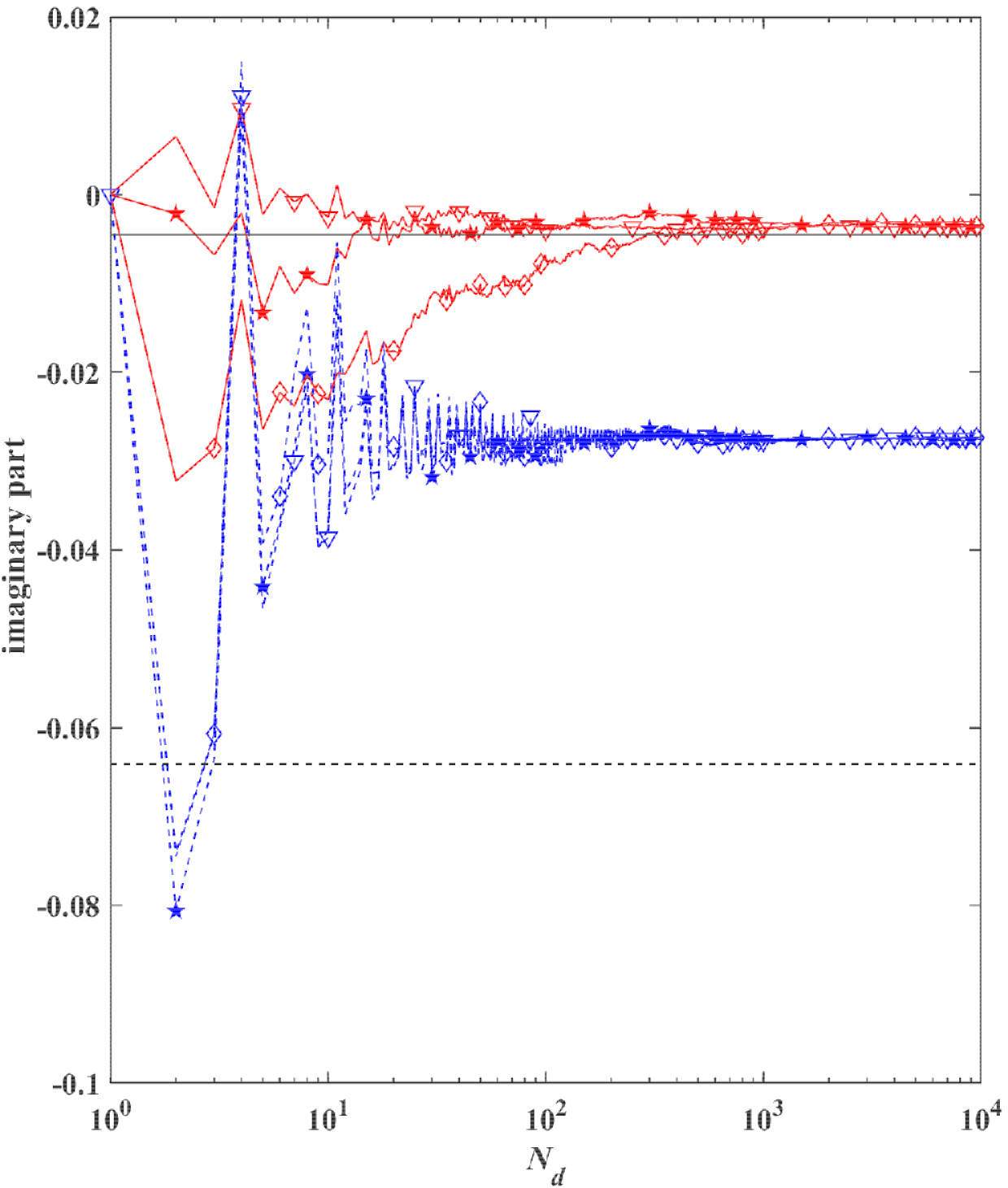}

\hspace*{1.0cm} (a) real part \hspace{2.0cm} (b) imaginary part

(1) estimate for $\phi_{u}(\mathbf{i}\omega_{0},\theta)$

\vspace{0.3cm}

\includegraphics[width=1.5in]{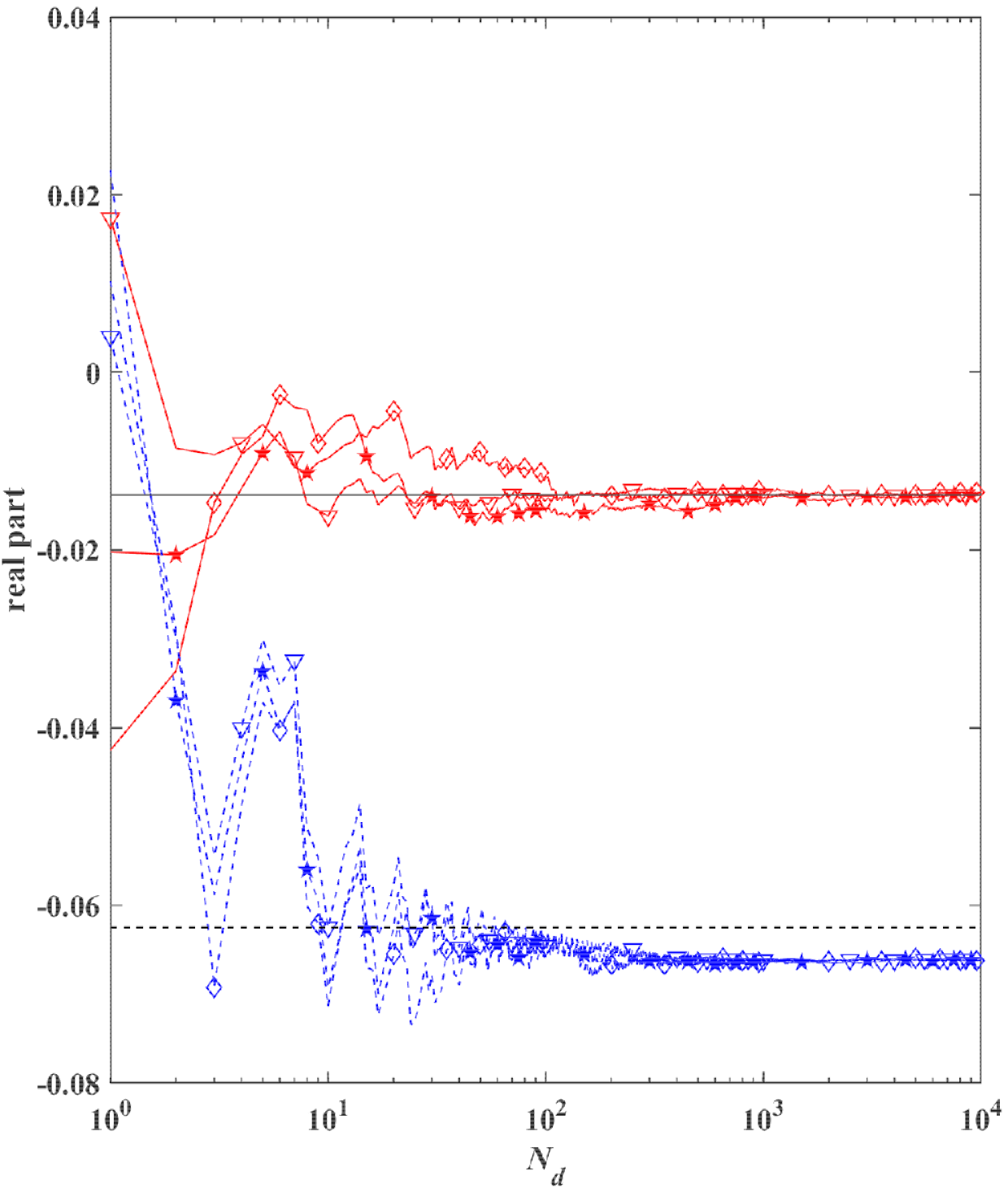}
\hspace{0.35cm}
\includegraphics[width=1.5in]{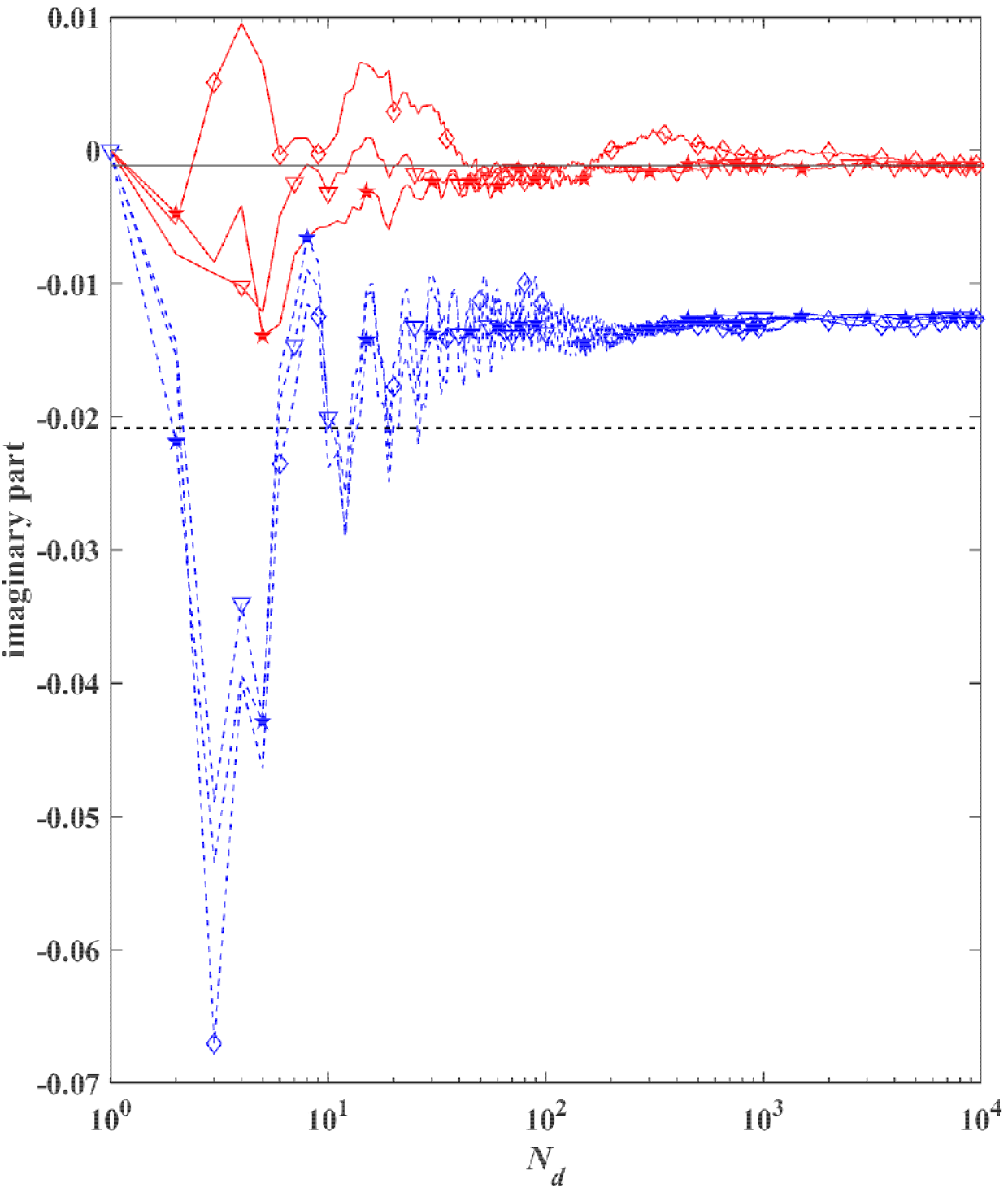}

\hspace*{1.0cm} (a) real part  \hspace{2.0cm} (b) imaginary part

(2) estimate for $\phi_{u}(\mathbf{i}\omega_{0} + \mathbf{i}\omega_{0},\theta)$

\vspace{0.3cm}
\includegraphics[width=1.5in]{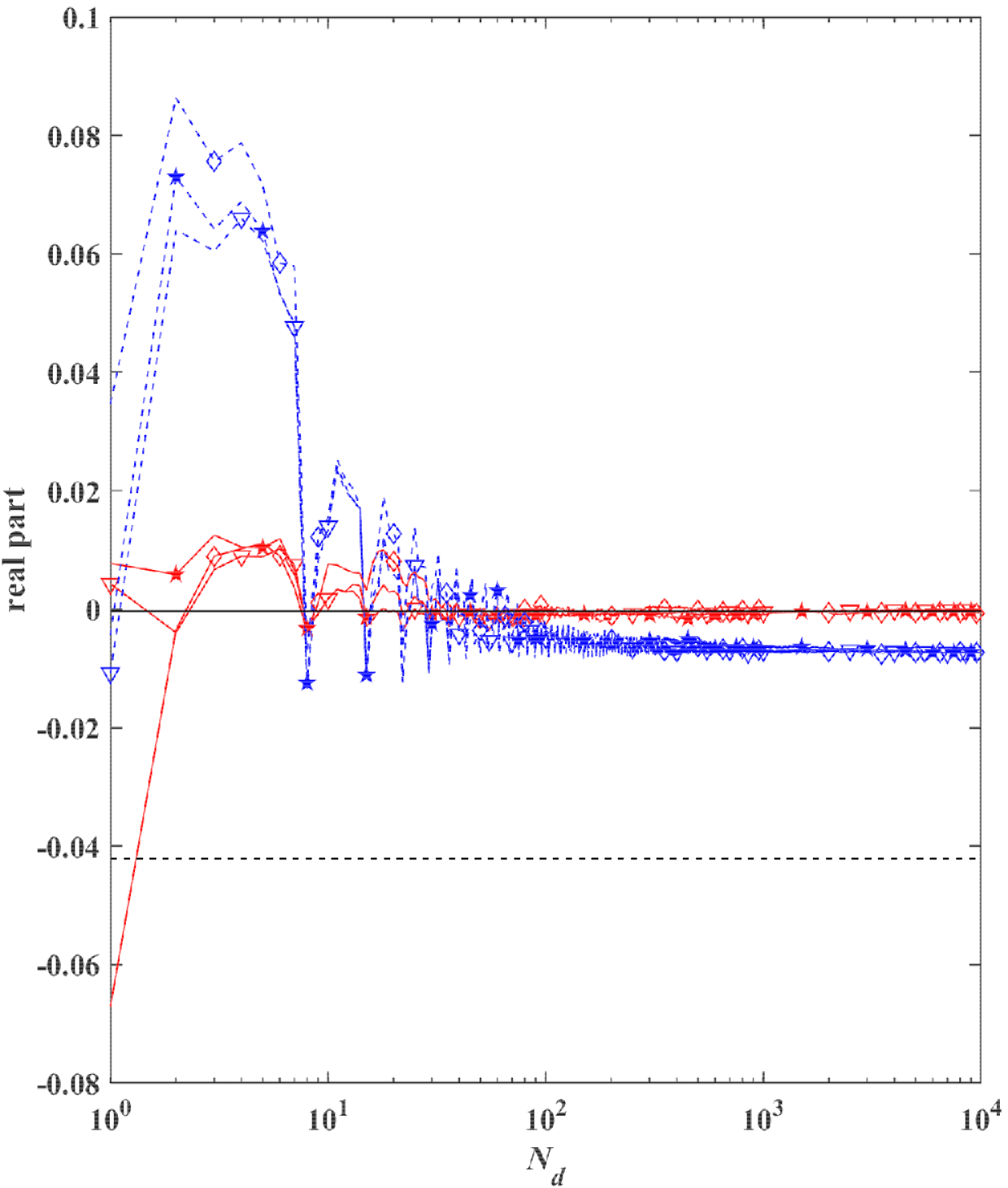}
\hspace{0.35cm}
\includegraphics[width=1.5in]{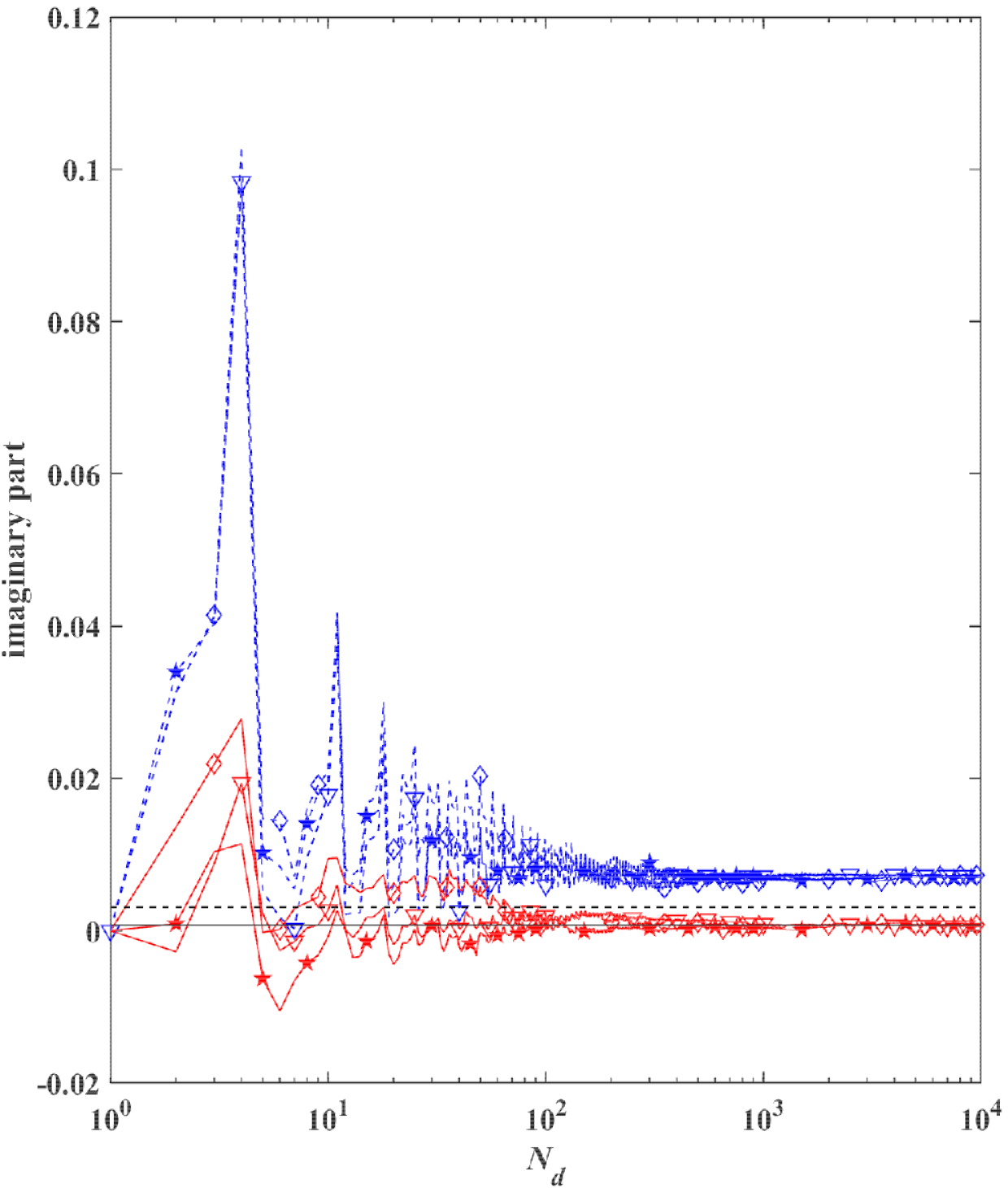}

\hspace*{1.0cm} (a) real part  \hspace{2.0cm} (b) imaginary part

(3) estimate for $\phi_{u}(\mathbf{i}\omega_{0}, \mathbf{i}\omega_{0}, \theta)$

\vspace{-0.4cm}\hspace*{5cm} \caption{Nonparametric Estimates with $\omega_{0} = 4.5 rad/s$. $-\!\!-$: actual value and its estimates for the 1st element; $-\:-$: actual value and its estimates for the 2nd element. $\nabla$: estimate with $\sigma=0.01$; $\star$: estimate with $\sigma=0.02$; $\bf\Diamond$: estimate with $\sigma=0.03$. }
\label{fig:2}
\vspace{-0.4cm}
\end{center}
\end{figure}

To simulate output measurements, a white noise is added to each voltage drop $v_{i}(t)$, $i=1,2$, that is independent of each other and has a normal distribution with expectation and standard deviation respectively being $0$ and $\sigma$. Several typical values are selected for the angular frequency $\omega_{0}$ and the standard deviation $\sigma$, as well as the data length $N_{d}$, in order to illustrate their influences on estimation accuracies, as well as influences of a nonparametric estimate on the estimation accuracy of a parametric estimate.

In numerical simulations, the circuit parameters are chosen as follows,
\begin{eqnarray*}
& &\hspace*{-1.0cm}
V_{th,1}=0.04V, \hspace{0.35cm} C_{1}=20F, \hspace{0.35cm} I_{s,1}=0.6A  \\
& &\hspace*{-1.0cm}
V_{th,2}=0.05V, \hspace{0.35cm} C_{2}=4F, \hspace{0.50cm} I_{s,2}=0.6A
\end{eqnarray*}

With these parameters, the factor $[C_{i}V_{th,i}( \sum_{j=1}^{k} \!\! s_{j}) + I_{s,i}]^{-1}$ with $i=1,2$, has the following explicit expression,
\begin{eqnarray}
& & \frac{1}{0.8 \sum_{j=1}^{k} \!\! s_{j} + 0.6} \hspace{0.15cm}{\rm when} \hspace{0.15cm}  i=1   \label{eqn:num-sim-4}    \\
& & \frac{1}{0.2 \sum_{j=1}^{k} \!\! s_{j} + 0.6} \hspace{0.15cm}{\rm when} \hspace{0.15cm}  i=2   \label{eqn:num-sim-5}
\end{eqnarray}
which clearly have different bandwidth if we regard $\sum_{j=1}^{k} \!\! s_{j}$ as a generalized Laplace variable.

\begin{figure}[t]
\vspace{-0.0cm}
\begin{center}
\includegraphics[width=3.0in]{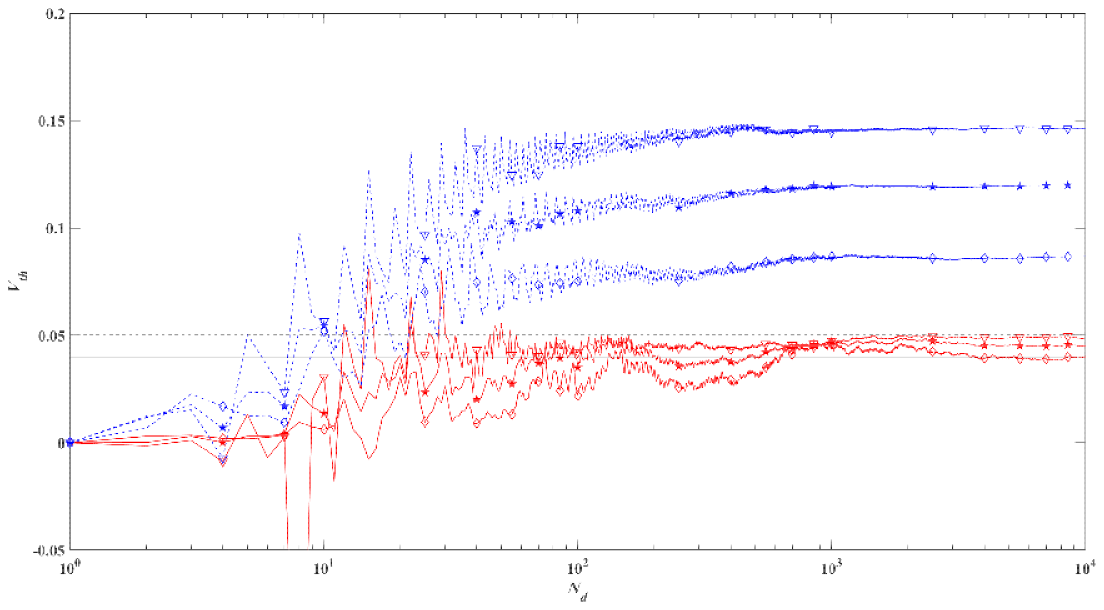}

(1) estimate with noise level $\sigma = 0.01$
\vspace{0.3cm}

\includegraphics[width=3.0in]{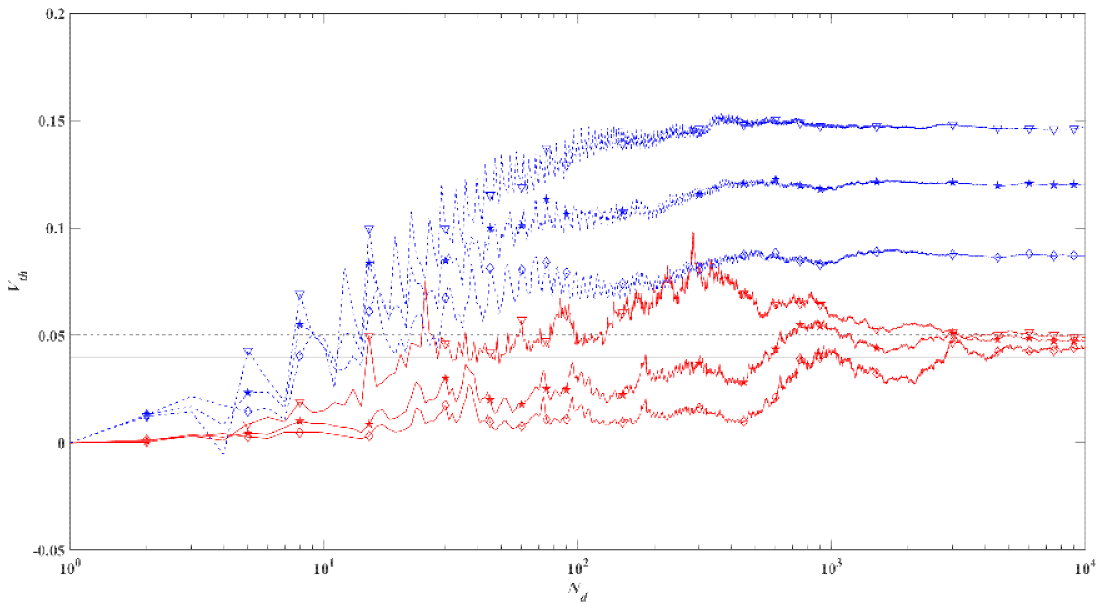}

(2) estimate with noise level $\sigma = 0.02$
\vspace{0.3cm}

\includegraphics[width=3.0in]{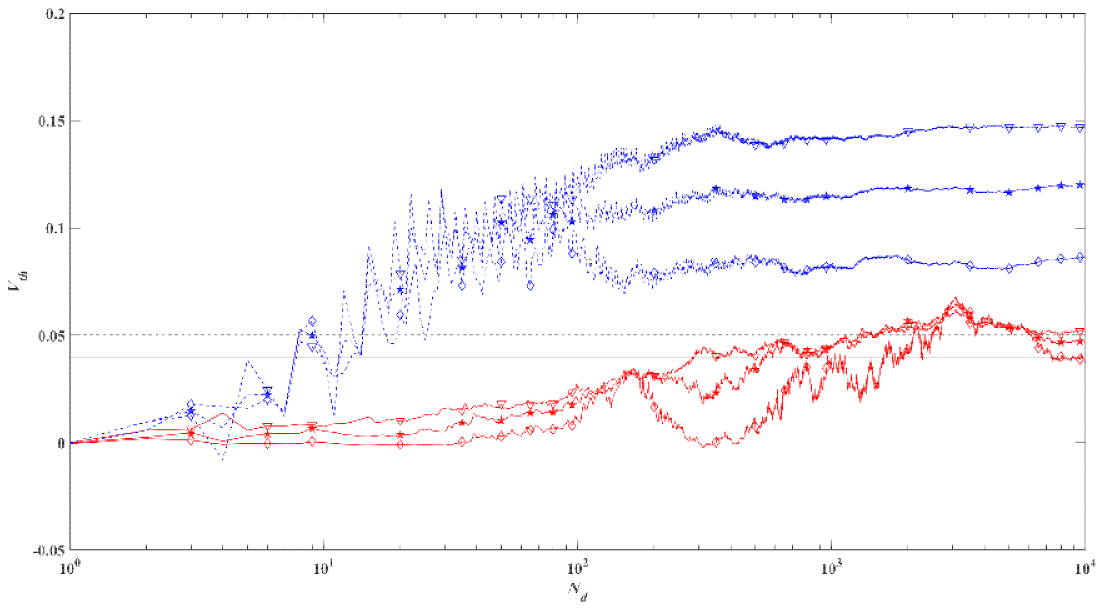}

(3) estimate with noise level $\sigma = 0.03$

\vspace{-0.4cm}\hspace*{5cm} \caption{Parametric Estimates with $\omega_{0} = 4.5 rad/s$. $-\!\!-$: actual value and its estimates for $V_{th,1}$; $-\:-$: actual value and its estimates for $V_{th,2}$. $\nabla$: estimate with $\widehat{\phi}_{u}(\mathbf{i}\omega_{0},\theta)$; $\star$: estimate with $\widehat{\phi}_{u}(\mathbf{i}\omega_{0},\theta)$ and $\widehat{\phi}_{u}(\mathbf{i}\omega_{0} + \mathbf{i}\omega_{0},\theta)$; $\bf\Diamond$: estimate with $\widehat{\phi}_{u}(\mathbf{i}\omega_{0},\theta)$, $\widehat{\phi}_{u}(\mathbf{i}\omega_{0} + \mathbf{i}\omega_{0},\theta)$ and $\widehat{\phi_{u}}(\mathbf{i}\omega_{0}, \mathbf{i}\omega_{0}, \theta)$.}
\label{fig:3}
\vspace{-0.4cm}
\end{center}
\end{figure}

Both nonparametric estimations for $\phi_{u}(\mathbf{i}\omega_{0},\theta)$,
$\phi_{u}(\mathbf{i}\omega_{0} + \mathbf{i}\omega_{0},\theta)$ and $\phi_{u}(\mathbf{i}\omega_{0}, \mathbf{i}\omega_{0}, \theta)$, and parametric estimations for $V_{th,1}$ and $V_{th,1}$, are performed. The parametric estimation is respectively based solely on the estimate of $\phi_{u}(\mathbf{i}\omega_{0},\theta)$,
based on the estimates of both $\phi_{u}(\mathbf{i}\omega_{0},\theta)$ and $\phi_{u}(\mathbf{i}\omega_{0} + \mathbf{i}\omega_{0},\theta)$, and based on the estimates of $\phi_{u}(\mathbf{i}\omega_{0},\theta)$, $\phi_{u}(\mathbf{i}\omega_{0} + \mathbf{i}\omega_{0},\theta)$ and $\phi_{u}(\mathbf{i}\omega_{0}, \mathbf{i}\omega_{0}, \theta)$, in order to investigate estimation accuracy improvements with the incorporation of information about high order harmonics in system response. Here with a little abuse of terminology, $\phi_{u}(1,\theta)$, $\phi_{u}(1, 1, \theta)$, etc. are expressed as $\phi_{u}(\mathbf{i}\omega_{0},\theta)$, $\phi_{u}(\mathbf{i}\omega_{0}, \mathbf{i}\omega_{0}, \theta)$, etc.  respectively, in order to clarify the dependence of these nonparametric estimates on the angular frequency $\omega_{0}$.

In these estimations, a nonparametric estimate is calculated using Equation (\ref{eqn:non-par-est-1}), while a parametric estimate is obtained through a least squares based data fitting in which each nonparametric estimate is treated equally, that is, with an equal weighting factor, using the techniques developed in \cite{Zhou2025} that are based on the LFT representations of the associated (generalized) TFMs.

\begin{figure}[t]
\vspace{-0.0cm}
\begin{center}

\includegraphics[width=1.5in]{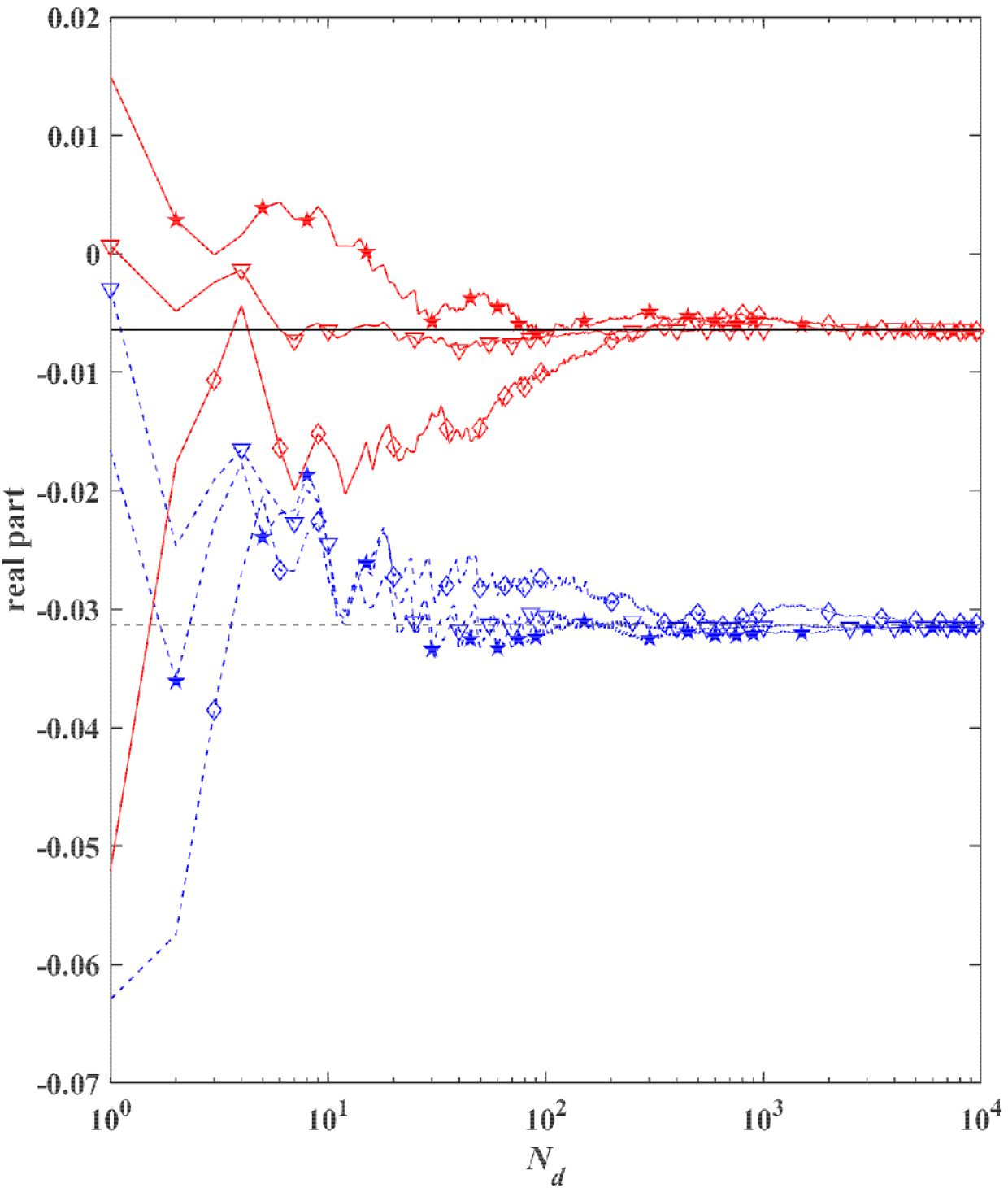}
\hspace{0.35cm}
\includegraphics[width=1.5in]{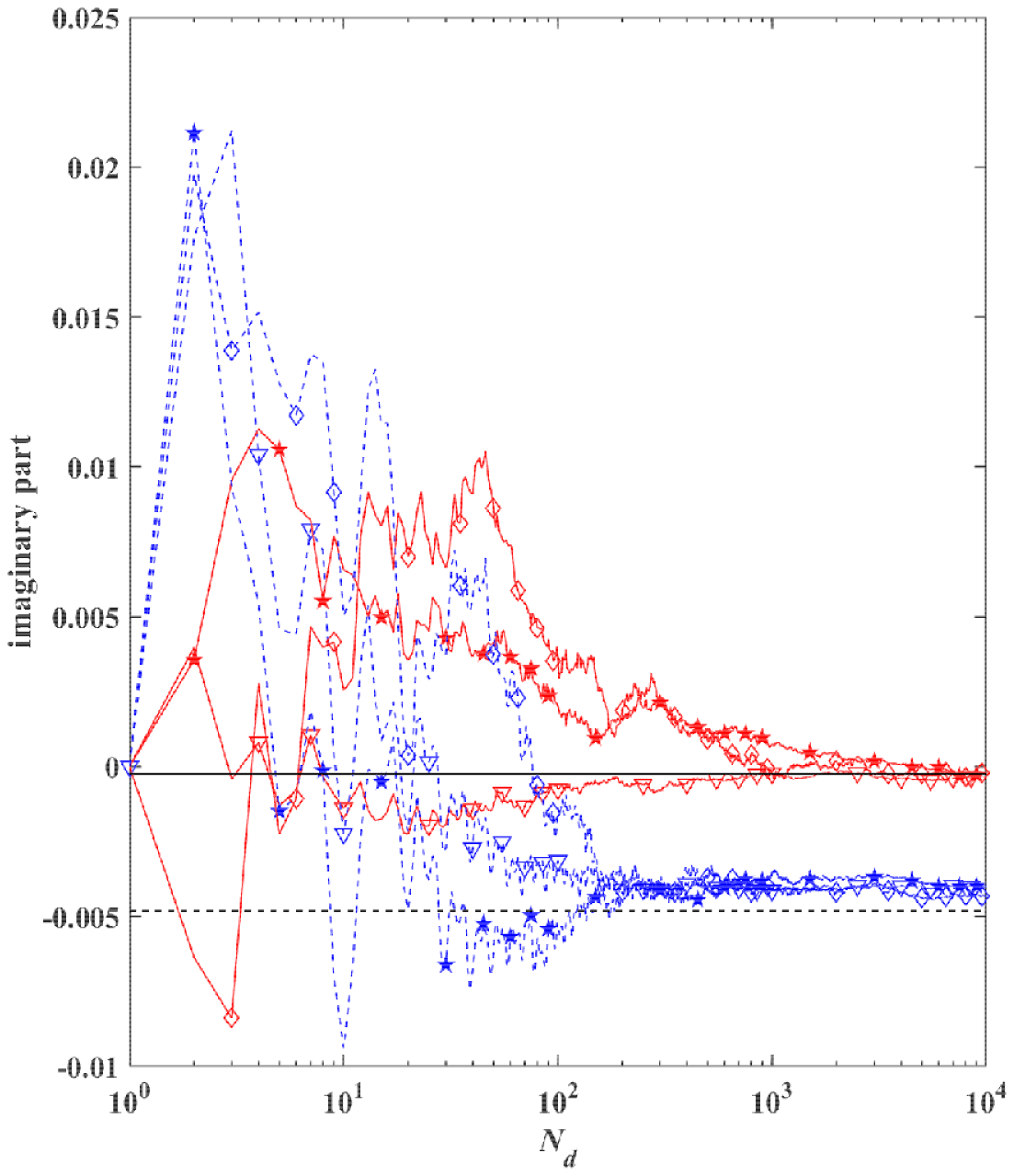}

\hspace*{1.0cm} (a) real part \hspace{2.0cm} (b) imaginary part

(1) estimate for $\phi_{u}(\mathbf{i}\omega_{0},\theta)$
\vspace{0.3cm}

\includegraphics[width=1.5in]{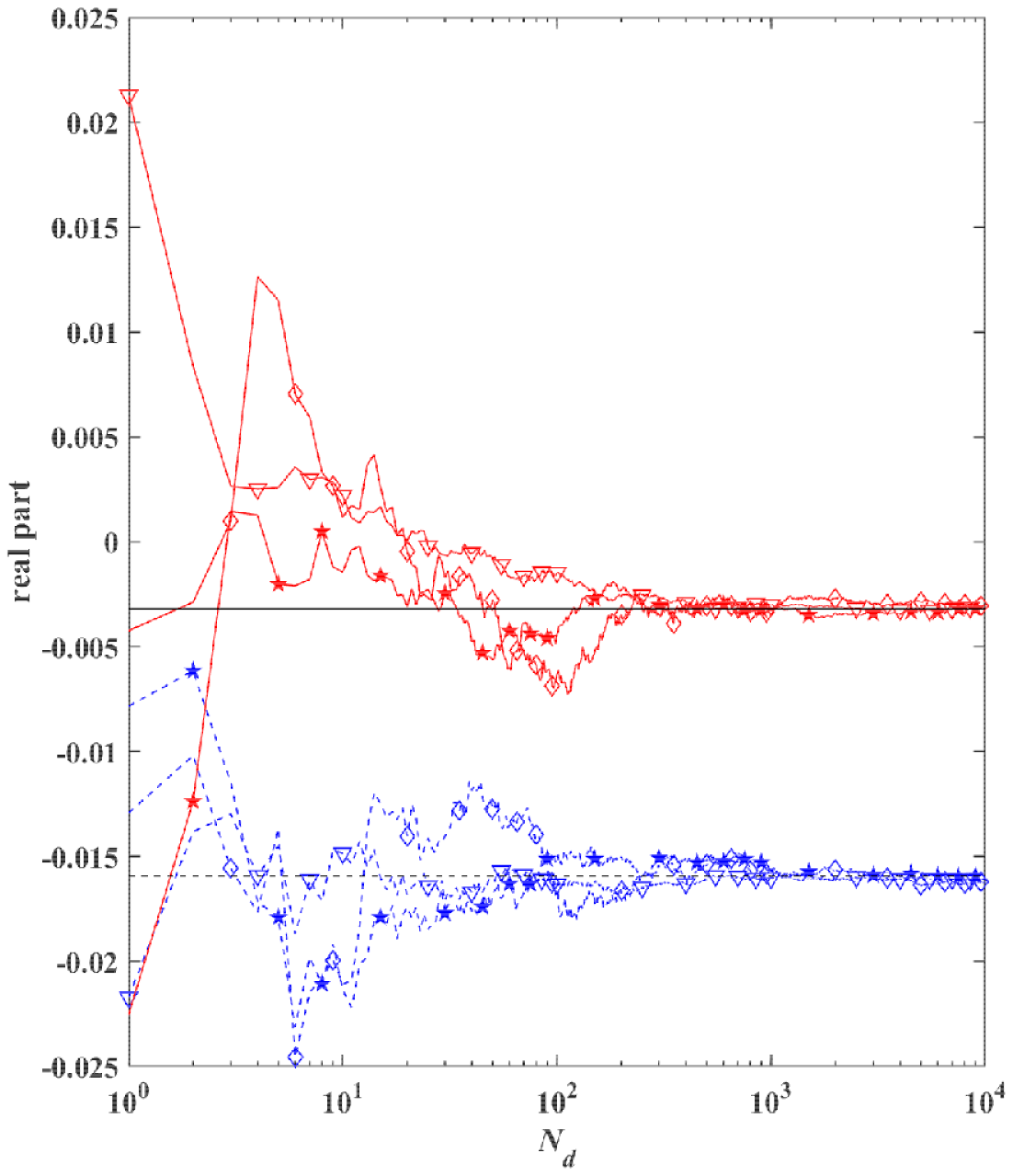}
\hspace{0.35cm}
\includegraphics[width=1.5in]{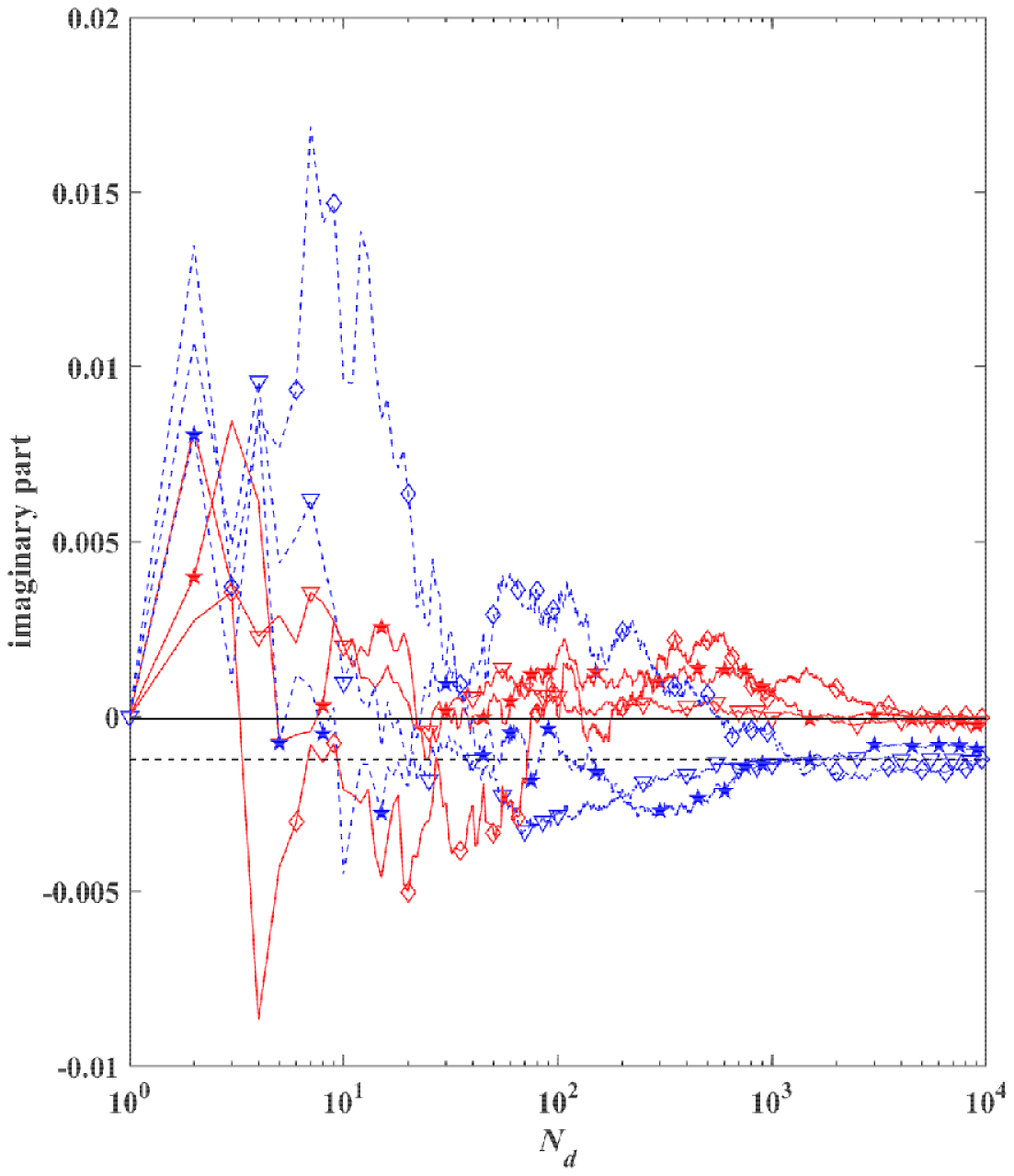}

\hspace*{1.0cm} (a) real part  \hspace{2.0cm} (b) imaginary part

(2) estimate for $\phi_{u}(\mathbf{i}\omega_{0} + \mathbf{i}\omega_{0},\theta)$
\vspace{0.3cm}

\includegraphics[width=1.5in]{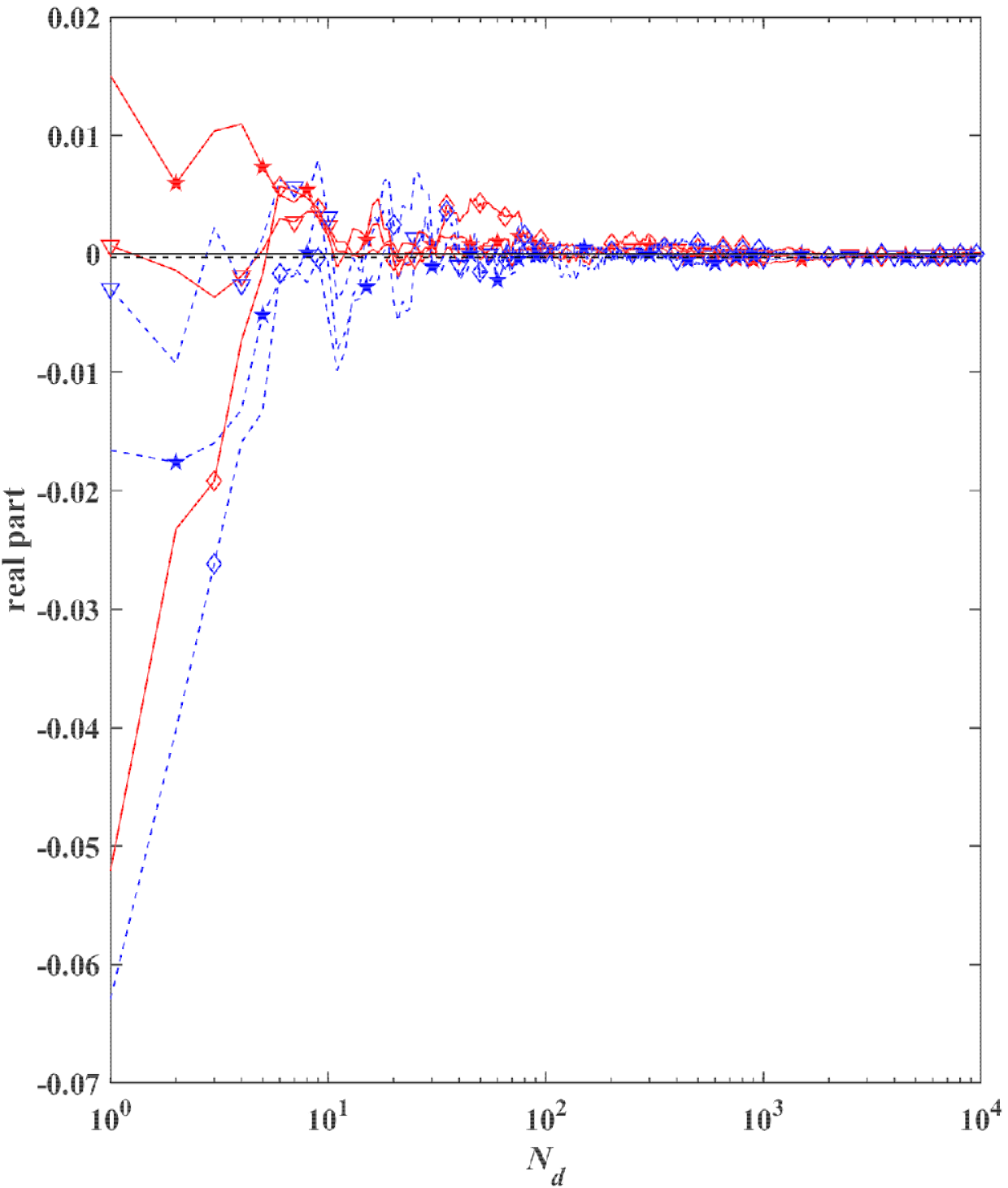}
\hspace{0.35cm}
\includegraphics[width=1.5in]{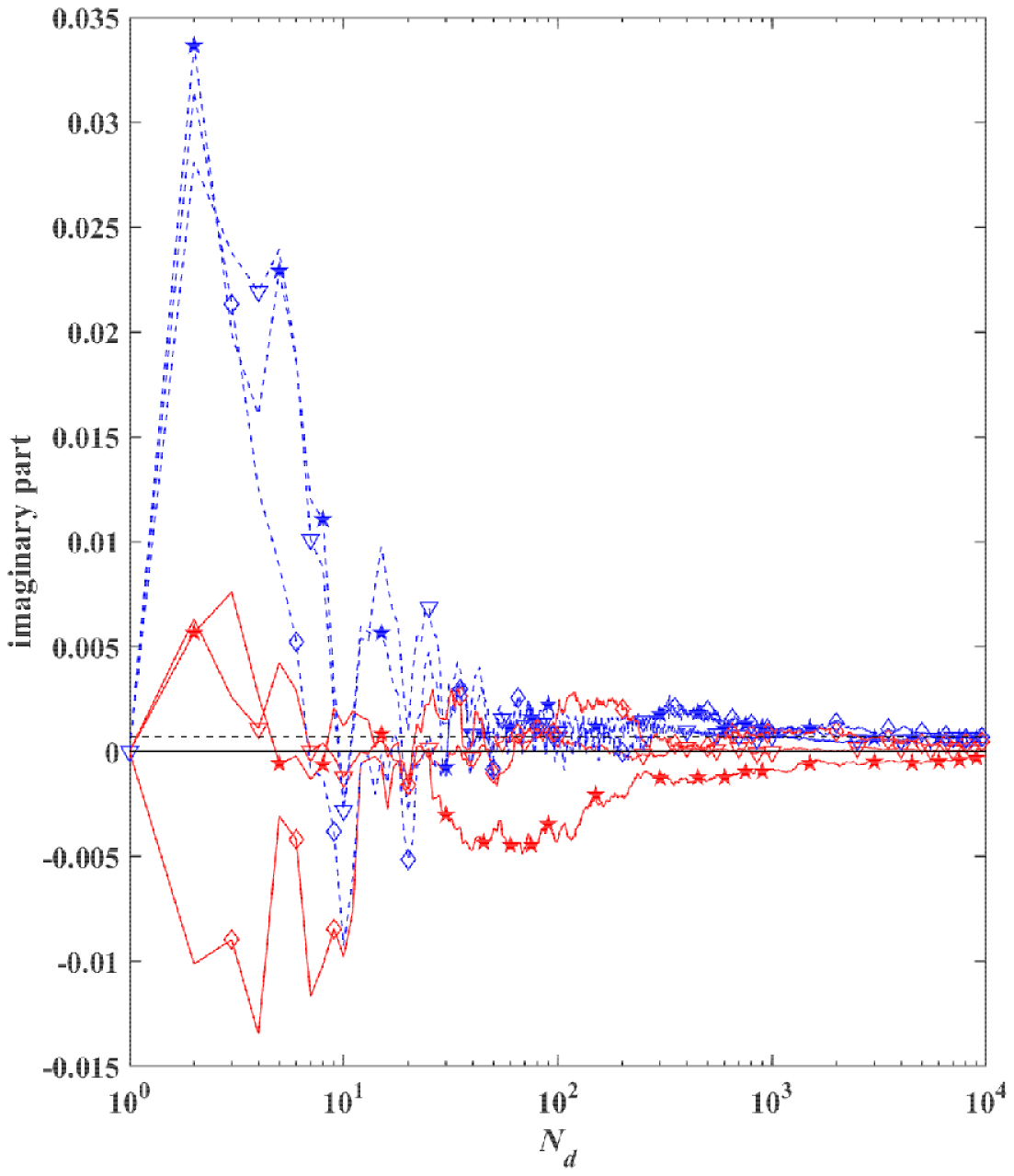}

\hspace*{1.0cm} (a) real part  \hspace{2.0cm} (b) imaginary part

(3) estimate for $\phi_{u}(\mathbf{i}\omega_{0}, \mathbf{i}\omega_{0}, \theta)$

\vspace{-0.4cm}\hspace*{5cm} \caption{Nonparametric Estimates with $\omega_{0} = 19.5 rad/s$. $-\!\!-$: actual value and its estimates for the 1st element; $-\:-$: actual value and its estimates for the 2nd element. $\nabla$: estimate with $\sigma=0.01$; $\star$: estimate with $\sigma=0.02$; $\bf\Diamond$: estimate with $\sigma=0.03$. }
\label{fig:4}
\vspace{-0.4cm}
\end{center}
\end{figure}

Some typical results are given in Figures 2-5 for nonparametric and parametric estimations\footnote[1]{In Figure 5, which gives results for parametric estimations with $\omega_{0} = 19.5 rad/s$, some curves have several discontinuous places. Rather than numerical instability, it is due to the display range selections. The ranges there are chosen for more clearly reflecting asymptotic properties of the computed estimates.}, with the data length $N_{d}$ increasing one by one from $1$ to $10^{4}$.

From these simulations, it is clear that all the estimates converge with the increment of experiment data length $N_{d}$. But some of them are approximately unbiased asymptotically, while the others do not have these properties. More specifically, when $\omega_{0} = 4.5 rad/s$, both the nonparametric estimates and the parametric estimates associated with $V_{th,1}$ are approximately unbiased asymptotically, but these properties are not shared with the $V_{th,2}$ associated nonparametric and parametric estimates. When the angular frequency $\omega_{0}$ is increased to $19.5 rad/s$, the asymptotic bias has been significantly reduced for both the nonparametric estimates and the parametric estimate associated with $V_{th,2}$, but convergence rate has been greatly reduced for estimates associated with both $V_{th,1}$ and $V_{th,2}$. These can be explained by that with the increment of the angular frequency of the input signal of the circuit, the signal to noise ratio of the measured circuit outputs is reduced, noting that for each associated (generalized) TFMs, its value decreases strictly monotonically in magnitude when this angular frequency increases, while the standard deviation of the measurement noise keeps unchanged. It can therefore be declared from Equations (\ref{eqn:resdec-1}) and (\ref{eqn:resdec-2}) that, the steady-state response of the circuit decreases in magnitude in a strictly monotonic way, and therefore the signal to noise ratio.

On the other hand, when $\omega_{0} = 4.5 rad/s$, $[\phi_{u}(\mathbf{i}\omega_{0},\theta)]_{1}$, $[\phi_{u}(\mathbf{i}\omega_{0} + \mathbf{i}\omega_{0},\theta)]_{1}$ and $[\phi_{u}(\mathbf{i}\omega_{0}, \mathbf{i}\omega_{0}, \theta)]_{1}$ are significantly greater in magnitude than the other elements in the steady-state response of $y_{1}(t)$ that are associated with the same angular frequency; but the elements associated respectively with $[\phi_{u}(\mathbf{i}\omega_{0},\theta)]_{2}$, $[\phi_{u}(\mathbf{i}\omega_{0} + \mathbf{i}\omega_{0},\theta)]_{2}$ and $[\phi_{u}(\mathbf{i}\omega_{0}, \mathbf{i}\omega_{0}, \theta)]_{2}$ are not very dominant, compared with other elements in the steady-state response of $y_{2}(t)$ that are associated with the same angular frequency. When the angular frequency $\omega_{0}$ is increased to $19.5 rad/s$, in the steady-state response of both $y_{1}(t)$ and $y_{2}(t)$, elements associated with these interpolation conditions become dominant. These dominance makes the approximation of Equation (\ref{eqn:non-par-est-1}) valid, and are consistent with the generalized TFMs of Equations (\ref{eqn:num-sim-1}) and(\ref{eqn:num-sim-2}), as well as the decaying ratios given by Equations (\ref{eqn:num-sim-4}) and (\ref{eqn:num-sim-5}).

\begin{figure}[t]
\vspace{-0.0cm}
\begin{center}
\includegraphics[width=3.0in]{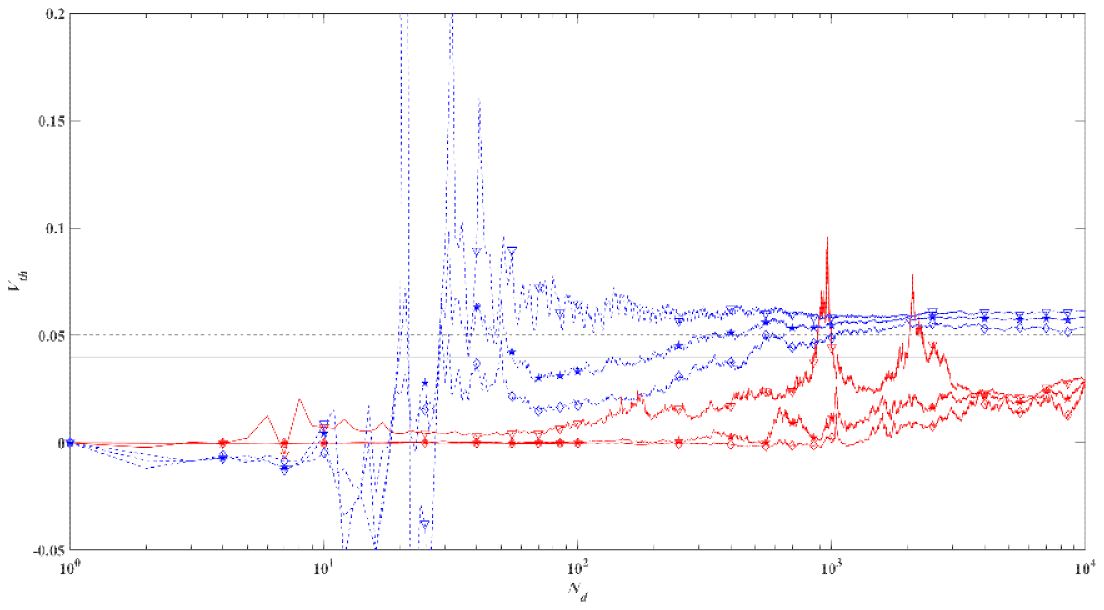}

(1) estimate with noise level $\sigma = 0.01$
\vspace{0.3cm}

\includegraphics[width=3.0in]{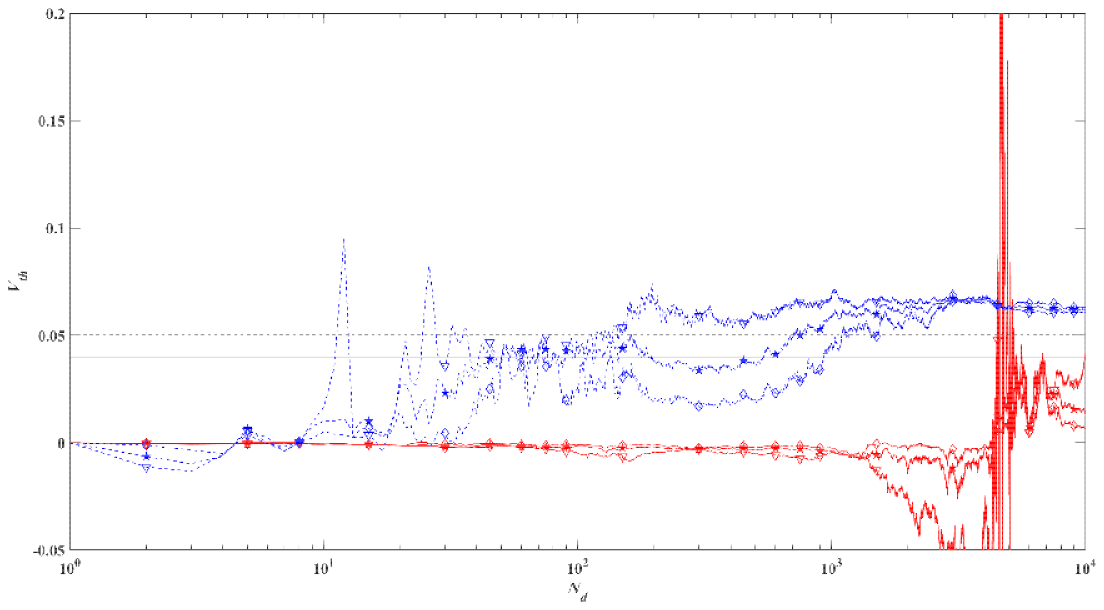}

(2) estimate with noise level $\sigma = 0.02$
\vspace{0.3cm}

\includegraphics[width=3.0in]{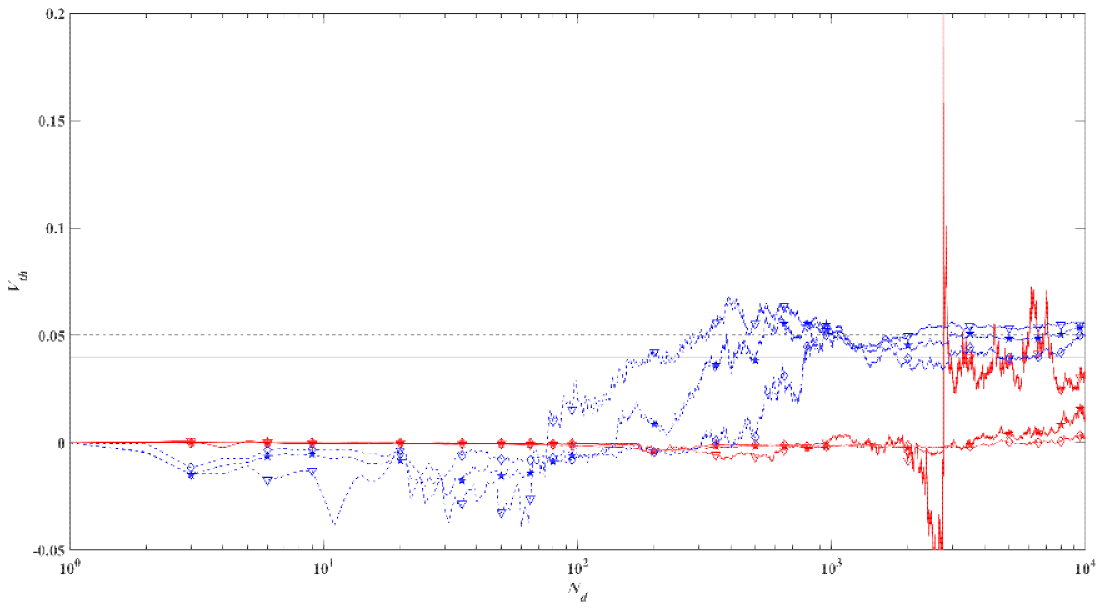}

(3) estimate with noise level $\sigma = 0.03$

\vspace{-0.4cm}\hspace*{5cm} \caption{Parametric Estimates with $\omega_{0} = 19.5 rad/s$. $-\!\!-$: actual value and its estimates for $V_{th,1}$; $-\:-$: actual value and its estimates for $V_{th,2}$. $\nabla$: estimate with $\widehat{\phi}_{u}(\mathbf{i}\omega_{0},\theta)$; $\star$: estimate with $\widehat{\phi}_{u}(\mathbf{i}\omega_{0},\theta)$ and $\widehat{\phi}_{u}(\mathbf{i}\omega_{0} + \mathbf{i}\omega_{0},\theta)$; $\bf\Diamond$: estimate with $\widehat{\phi}_{u}(\mathbf{i}\omega_{0},\theta)$, $\widehat{\phi}_{u}(\mathbf{i}\omega_{0} + \mathbf{i}\omega_{0},\theta)$ and $\widehat{\phi_{u}}(\mathbf{i}\omega_{0}, \mathbf{i}\omega_{0}, \theta)$.}
\label{fig:5}
\vspace{-0.4cm}
\end{center}
\end{figure}

Another observation from these simulation results is that when $\omega_{0} = 4.5 rad/s$ and the experimental data length is sufficiently large, estimation accuracy increases monotonically for both $V_{th,1}$ and $V_{th,2}$ with the addition of a new nonparametric estimate, that is, an estimate for $[\phi_{u}(\mathbf{i}\omega_{0} + \mathbf{i}\omega_{0},\theta)]_{i}$ or $[\phi_{u}(\mathbf{i}\omega_{0}, \mathbf{i}\omega_{0}, \theta)]_{i}$ with $i=1,2$. However, this is not the case when $\omega_{0} = 19.5 rad/s$. More specifically, even when the experimental data length $N_{d}$ is close to $10^{4}$, although the estimation accuracy for $V_{th,2}$ with $\sigma=0.01$ increases monotonically with the addition of a new nonparametric estimate, its estimation accuracy decreases with the introduction of a new nonparametric estimate when $\sigma=0.02$, and becomes unclear when $\sigma=0.03$. This is not surprising, noting that in the parametric estimations, each of the nonparametric estimates is fitted with an equal weight, implicitly assuming that they have the same estimation accuracy. As magnitudes of the involved tangential interpolation conditions are generally different, and they are estimated using the same simulated experimental data or experimental data with the same simulation settings, their estimation accuracies are different in general. These mean that in order to efficiently utilize information contained in a nonparametric estimate, an appropriate weighting factor must be introduced. As a matter of fact, when some weighting factors are introduced into this numerical example and are suitably adjusted, accuracy can always be improved when a nonparametric estimate is added for the parametric estimation. However, further efforts are needed to develop a systematic method for this weighting factor selection.

\section{Conclusions}\label{sec:conclusions}

Estimation is studied for SIPs of an NDS whose subsystems are described by a continuous-time QBTI model. No restrictions are put on the sampling rate. Some explicit formulas have been obtained for the harmonics of the NDS time-domain response. A linear dependence relation has been established between NDS steady-state response and the values of its TFM and generalized TFMs at several particular locations. These TFMs are completely determined by its system matrices, and depend on the NDS SIPs though an LFT. An estimate is derived respectively for the tangential interpolation of these TFMs, and the NDS SIP vector.

Further efforts include how to efficiently incorporate transient responses of the NDS into an estimation for its SIPs, as well as how to remove the uniform sampling constraint. It is also interesting to see how to find appropriate locations and directions for estimating a tangential interpolation condition of an associated (generalized) TFM, that lead to a high accuracy estimate of the SIP vector with a low computational complexity; as well as how to incorporate estimates together that are for different tangential interpolation conditions with an appropriate weighting factor, such that their information about the SIP vector is efficiently utilized.

\small
\vspace{0.25cm}

\renewcommand{\labelenumi}{\rm\bf A\arabic{enumi})}

\renewcommand{\theequation}{A\arabic{equation}}
\setcounter{equation}{0}

\small
\appendices
\section*{Appendix I. Proof of Some Technical Results}

\hspace*{-0.40cm}{\rm\bf Proof of Theorem \ref{theo:1}.}
Recall that in the frequency domain, existence of an impulse mode in a descriptor system is equivalent to that the inverse of $s E-A$ is not strictly proper even it is regular \cite{Dai1989,Duan2010}. It can therefore be declared that when Assumptions \ref{assum:1}-\ref{assum:10} are simultaneously satisfied by the NDS $\mathbf{\Sigma}_{p}$ and the assembly PSGS $\mathbf{\Sigma}_{s}$, the matrix $sE-A$ is invertible, and its inverse is strictly proper. These mean that there exist constant matrices $P_{i} \in \mathbb{C}^{m_{x}\times m_{x}}$, $i=1,2,\cdots,m_{x}$, such that
\begin{equation}
(sE-A)^{-1} = \sum_{i=1}^{m_{x}}\frac{P_{i}}{s-\lambda_{p,i}}
\label{app:1-1}
\end{equation}

On the other hand, from the state space model of the assembly PSGS $\mathbf{\Sigma}_{s}$ and the definition of the vectors  $\psi_{u}(i)|_{i=1}^{m_{\xi}}$, it can be straightforwardly shown that
\begin{equation}
\hspace*{-0.2cm}
u(t) = \Pi T_{s} diag\left\{e^{\lambda_{s,i}t}|_{i=1}^{m_{\xi}} \right\}T_{s}^{-1}\xi(0)
= \sum_{i=1}^{m_{\xi}}e^{\lambda_{s,i}t} \psi_{u}(i)
\label{app:1-5}
\end{equation}

Using the symbols of Lemmas \ref{lemma:2} and \ref{lemma:3}, define vectors $\psi_{p}(i)|_{i=1}^{m_{x}}$ and $\psi_{s}(i)|_{i=1}^{m_{\xi}}$ respectively as
\begin{eqnarray*}
& & \hspace*{-1.0cm} \psi_{p}(i) = P_{i}E\left[x_{1}(0) \!-\! X\xi(0)\right]   \\
& & \hspace*{-1.0cm} \psi_{s}(i) = (\lambda_{s,i}E-A)^{-1} \!B \psi_{u}(i)
\end{eqnarray*}
Then it can be declared from these two lemmas that if the linear part of the NDS $\mathbf{\Sigma}_{p}$ does not have a generalized eigenvalue that is equal to an eigenvalue of the assembly PSGS $\mathbf{\Sigma}_{s}$, then there exists a constant matrix $X \in \mathbb{R}^{m_{x}\times m_{\xi}}$, such that the definition of the aforementioned vectors $\psi_{p}(i)|_{i=1}^{m_{x}}$ is well-posed, and
\begin{equation}
\hspace*{-0.8cm}
x_{1}(t) = \sum_{i=1}^{m_{x}} \!e^{\lambda_{p,i}t} \psi_{p}(i)  \!+\!
\sum_{i=1}^{m_{\xi}}e^{\lambda_{s,i}t} \psi_{s}(i)
\label{app:1-4}
\end{equation}
meaning that the conclusion is valid for $k=1$.

With these expressions for $x_{1}(t)$ and $u(t)$, the following equalities can be established directly from properties of matrix Kronecker products,
\begin{eqnarray}
& & \hspace*{-0.6cm}
x_{1}(t) \otimes x_{1}(t) = \sum_{i_{1}=1}^{m_{x}}\sum_{i_{2}=1}^{m_{x}} \!e^{(\lambda_{p,i_{1}}+\lambda_{p,i_{2}})t} \psi_{p}(i_{1})\otimes \psi_{p}(i_{2})  \!+ \nonumber\\
& & \hspace*{0.2cm}
\sum_{i_{1}=1}^{m_{x}} \!\sum_{i_{2}=1}^{m_{\xi}} \!\!e^{(\lambda_{p,i_{1}} \!+\! \lambda_{s,i_{2}})t} \left[\psi_{p}(i_{1}) \!\otimes\! \psi_{s}(i_{2}) \!+\! \psi_{s}(i_{2}) \!\otimes\! \psi_{p}(i_{1})\right]  \!+ \nonumber\\
& & \hspace*{0.2cm}
\sum_{i_{1}=1}^{m_{\xi}}\sum_{i_{2}=1}^{m_{\xi}}e^{(\lambda_{s,i_{1}}+\lambda_{s,i_{2}})t} \psi_{s}(i_{1})\otimes \psi_{s}(i_{2})
\label{app:1-6}\\
& & \hspace*{-0.6cm}
x_{1}(t) \otimes u(t) = \sum_{i_{1}=1}^{m_{x}}\sum_{i_{2}=1}^{m_{\xi}} \!e^{(\lambda_{p,i_{1}}+\lambda_{s,i_{2}})t} \psi_{p}(i_{1})\otimes \psi_{u}(i_{2})  \!+ \nonumber\\
& & \hspace*{0.6cm}
\sum_{i_{1}=1}^{m_{\xi}}\sum_{i_{2}=1}^{m_{\xi}} \!e^{(\lambda_{s,i_{1}}+\lambda_{s,i_{2}})t} \psi_{s}(i_{1}) \!\otimes\! \psi_{u}(i_{2})
\label{app:1-7}
\end{eqnarray}
Therefore,
\begin{eqnarray}
& & \hspace*{-0.6cm}
\Gamma_{x}\left[x_{1}(t) \otimes x_{1}(t)\right] + \Gamma_{u}\left[x_{1}(t) \otimes u(t)\right] \nonumber \\
& & \hspace*{-1.0cm}= \!\! \sum_{i_{1}=1}^{m_{x}}\sum_{i_{2}=1}^{m_{x}} \!e^{(\lambda_{p,i_{1}}+\lambda_{p,i_{2}})t} \overline{\psi}_{p}(i_{l}\!|_{l=1}^{2})  \!+ \nonumber\\
& & \hspace*{-0.8cm}
\sum_{i_{1}=1}^{m_{x}}\sum_{i_{2}=1}^{m_{\xi}} \!e^{(\lambda_{p,i_{1}}+\lambda_{s,i_{2}})t} \overline{\psi}_{p,s}(i_{l}\!|_{l=1}^{2})\!+ \nonumber\\
& & \hspace*{-0.8cm}
\sum_{i_{1}=1}^{m_{\xi}}\sum_{i_{2}=1}^{m_{\xi}}e^{(\lambda_{s,i_{1}}+\lambda_{s,i_{2}})t} \overline{\psi}_{s}(i_{l}\!|_{l=1}^{2})
\label{app:1-8}
\end{eqnarray}
in which
\begin{eqnarray*}
& & \hspace*{-0.8cm}
\overline{\psi}_{p}(i_{l}\!|_{l=1}^{2}) \!= \! \Gamma_{x}  \! \!\left[ \psi_{p}(i_{1})\otimes \psi_{p}(i_{2})\right]  \\
& & \hspace*{-0.8cm}
\overline{\psi}_{p,s}(i_{l}\!|_{l=1}^{2})  \!= \! \Gamma_{x}  \! \!\left[\psi_{p}(i_{1}) \!\otimes\! \psi_{s}(i_{2})  \!+ \! \psi_{s}(i_{2}) \!\otimes\! \psi_{p}(i_{1})\right]  \!+ \! \\
& & \hspace*{0.8cm} \Gamma_{u}\left[\psi_{p}(i_{1}) \!\otimes\! \psi_{u}(i_{2}) \right]  \\
& & \hspace*{-0.8cm}
\overline{\psi}_{s}(i_{l}\!|_{l=1}^{2})  \!= \! \Gamma_{x}  \! \!\left[\psi_{s}(i_{1})  \!\otimes \! \psi_{s}(i_{2})\right]  \!+ \!  \Gamma_{u} \! \!\left[\psi_{s}(i_{1})  \!\otimes \! \psi_{u}(i_{2}) \right]
\end{eqnarray*}

On the other hand, from Lemma \ref{lemma:2}, we have that
\begin{displaymath}
\hspace*{-0.2cm}
E \dot{x}_{2}(t) \!=\! Ax_{2}(t) \!+\! \Gamma_{x} \!\left[x_{1}(t) \!\otimes\! x_{1}(t)\right] \!+\!
           \Gamma_{u}\!\left[x_{1}(t) \!\otimes\! u(t)\right]
\end{displaymath}
Substitute Equation (\ref{app:1-8}) into this equation, and take Laplace transform on both of its left and right sides. Then the following equality is obtained
\begin{eqnarray}
& & \hspace*{-1.0cm}
E \left[{x}_{2}(s)-{x}_{2}(0)\right] \!=\! Ax_{2}(s) \!+\! \sum_{i_{1}=1}^{m_{x}}\sum_{i_{2}=1}^{m_{x}} \!\frac{\overline{\psi}_{p}(i_{l}\!|_{l=1}^{2})}{s-(\lambda_{p,i_{1}}+\lambda_{p,i_{2}})}  \!+ \nonumber\\
& & \hspace*{-0.6cm}
\sum_{i_{1}=1}^{m_{x}} \!\sum_{i_{2}=1}^{m_{\xi}} \!\! \frac{\overline{\psi}_{p,s}(i_{l}\!|_{l=1}^{2})}{s \!-\! (\lambda_{p,i_{1}} \!+\! \lambda_{s,i_{2}})} \!+\!
\sum_{i_{1}=1}^{m_{\xi}} \!\sum_{i_{2}=1}^{m_{\xi}} \!\!\frac{\overline{\psi}_{s}(i_{l}\!|_{l=1}^{2})}{s \!-\! (\lambda_{s,i_{1}} \!+\! \lambda_{s,i_{2}})}
\label{app:1-9}
\end{eqnarray}
Therefore
\begin{eqnarray}
& & \hspace*{-1.0cm}
{x}_{2}(s) \!=\! (sE \!\!-\!\! A)^{\!-\!1}\!{x}_{2}(0) \!+\!\!
\sum_{i_{1}=1}^{m_{\xi}} \!\sum_{i_{2}=1}^{m_{\xi}} \!\!\frac{(sE \!\!-\!\! A)^{\!-\!1}\overline{\psi}_{s}(i_{l}\!|_{l=1}^{2})}{s \!-\! (\lambda_{s,i_{1}} \!+\! \lambda_{s,i_{2}})}   \!+\! \sum_{i_{1}=1}^{m_{x}}
\nonumber\\
& & \hspace*{-0.2cm}
 \!\!\!\!\left( \!\sum_{i_{2}=1}^{m_{x}} \!\!\!\frac{(sE \!\!-\!\! A)^{\!-\!1}\overline{\psi}_{p}(i_{l}\!|_{l=1}^{2}\!)}{s-(\lambda_{p,i_{1}}+\lambda_{p,i_{2}})} \!\!+\!\!\!
\sum_{i_{2}=1}^{m_{\xi}} \!\!\!\! \frac{{(sE \!\!-\!\! A)^{\!-\!1}\overline\psi}_{p,s}(i_{l}\!|_{l=1}^{2}\!)}{s \!-\! (\lambda_{p,i_{1}} \!+\! \lambda_{s,i_{2}})} \!\!\! \right)
\label{app:1-10}
\end{eqnarray}

Define vectors $\psi_{p}(i)|_{i=1}^{m_{x}}$, $\psi_{p,s}^{[2,1]}(i_{l}|_{l=1}^{2})|_{i_{1},i_{2}=1}^{i_{1}=m_{x},i_{2}=m_{\xi}}$, $\psi_{p,s}^{[2,2]}(i_{l}|_{l=1}^{2})|_{i_{1},i_{2}=1}^{m_{x}}$ and $\psi_{s}(i_{l}|_{l=1}^{2})|_{i_{1},i_{2}=1}^{m_{\xi}}$ respectively as
\begin{eqnarray*}
& & \hspace*{-1.0cm}
\psi_{p}(i) = P_{i}\left\{{x}_{2}(0) \!+\!
\sum_{i_{1}=1}^{m_{\xi}} \!\sum_{i_{2}=1}^{m_{\xi}} \!\!\frac{\overline{\psi}_{s}(i_{l}\!|_{l=1}^{2})}{ \lambda_{p,i}\!-\! (\lambda_{s,i_{1}} \!+\! \lambda_{s,i_{2}})}   + \right.  \\
& & \hspace*{-0.8cm} \left.
\sum_{i_{1}=1}^{m_{x}} \!\!\left( \!\sum_{i_{2}=1}^{m_{x}} \!\!\frac{\overline{\psi}_{p}(i_{l}\!|_{l=1}^{2})}{\lambda_{p,i} \!-\! (\lambda_{p,i_{1}} \!+\! \lambda_{p,i_{2}})} \!+\!\!
\sum_{i_{2}=1}^{m_{\xi}} \!\! \frac{\overline{\psi}_{p,s}(i_{l}\!|_{l=1}^{2})}{\lambda_{p,i} \!-\! (\lambda_{p,i_{1}} \!+\! \lambda_{s,i_{2}})} \!\!\! \right)\!\!\!\right\} \\
& & \hspace*{-1.0cm}
\psi_{p,s}^{[2,1]}(i_{l}|_{l=1}^{2}) = \left[(\lambda_{p,i_{1}}\!+\! \lambda_{s,i_{2}})E \!-\! A\right]^{\!-\!1}\overline{\psi}_{p,s}(i_{l}\!|_{l=1}^{2})   \\
& & \hspace*{-1.0cm}
\psi_{p,s}^{[2,2]}(i_{l}|_{l=1}^{2}) = \left[(\lambda_{p,i_{1}}\!+\! \lambda_{p,i_{2}})E \!-\! A\right]^{\!-\!1}\overline{\psi}_{p}(i_{l}\!|_{l=1}^{2})   \\
& & \hspace*{-1.0cm}
\psi_{s}(i_{l}|_{l=1}^{2}) = \left[(\lambda_{s,i_{1}}\!+\! \lambda_{s,i_{2}})E \!-\! A\right]^{\!-\!1}\overline{\psi}_{s}(i_{l}\!|_{l=1}^{2})
\end{eqnarray*}
Then, on the basis of Equations (\ref{app:1-1}) and (\ref{app:1-11}), as well as linearity properties of the Laplace transformation, the following expression for $x_{2}(t)$ can be established, taking inverse Laplace transformation on both sides of Equation (\ref{app:1-10}),
\begin{eqnarray}
& & \hspace*{-1.0cm}
x_{2}(t) \!=\!\! \sum_{i_{1}=1}^{m_{x}} \!e^{\lambda_{p,i_{1}}t} \!\!\left( \!\! \psi_{p}(i_{1}) \!+\!\!\left\{\!\!
\sum_{i_{2}=1}^{m_{x}}
\left[ \! e^{\lambda_{p,i_{2}}t} \psi_{p,s}^{[2,2]}(i_{l}\!|_{l=1}^{2}) \right]  \right.\right. \nonumber \\
& & \hspace*{2.2cm}
\left.\left.\sum_{i_{2}=1}^{m_{\xi}} \!\left[ \! e^{\lambda_{s,i_{2}}t} \psi_{p,s}^{[2,1]}(i_{l}\!|_{l=1}^{2}) \right] \!\!\right\}\!\!\right) \!+ \nonumber \\
& & \hspace*{-0.2cm}
\sum_{i_{1}=1}^{m_{\xi}}\sum_{i_{2}=1}^{m_{\xi}}e^{\lambda_{s}(i_{l}|_{l=1}^{2})t}\psi_{s}(i_{l}|_{l=1}^{2})
\label{app:1-11}
\end{eqnarray}

From the definition of $\lambda_{p,s}^{[l,q]}(i_{h}|_{h=1}^{l})$, it is obvious that $\lambda_{p,s}^{[2,2]}(i_{l}|_{l=1}^{2}) = \lambda_{p,i_{2}}$, while $\lambda_{p,s}^{[2,1]}(i_{l}|_{l=1}^{2}) = \!\lambda_{s,i_{2}}$ for every admissible pair of $i_{1}$ and $i_{2}$, meaning that the conclusions are valid for $k=2$.

Now assume that the expression of Equation (\ref{eqn:theo:1-2}) is valid for each $k=1,2,\cdots,m$. Then by the same token adopted in the proof for the case with $k=2$, it can be proved that this expression is also valid for $k=m+1$. The details are omitted due to its straightforwardness and lengthy equation expressions. This completes the proof. \hspace{\fill}$\Diamond$

\hspace*{-0.40cm}{\rm\bf Proof of Corollary \ref{coro:1}.}
From Lemma \ref{lemma:2} and Theorem \ref{theo:1}, we have straightforwardly that for each $k\geq 2$,
\begin{eqnarray}
& & \hspace*{-1.0cm} E \dot{x}_{k}(t) \!\!=\!\! Ax_{k}(t) \!+\! \Gamma_{\!x} \!\!\sum_{l=1}^{k \!-\!1}\!\!\left(\!\left[x_{l,t}(t) \!+\! x_{l,s}(t)\right] \!\!\otimes\!\! \left[x_{k\!-\!l,t}(t) \!+\! x_{k\!-\!l,s}(t)\right]\!\right) \!+\nonumber\\
& & \hspace*{2.8cm}
\Gamma_{u}\!\left(\left[x_{k \!-\! 1,t}(t) \!+\! x_{k \!-\! 1,s}(t)\right] \!\otimes\! u(t)\right)
\nonumber\\
& & \hspace*{-0.8cm} =\!\! Ax_{k\!}(t) \!+\! \Gamma_{\!x} \!\!\sum_{l=1}^{k \!-\!1}\!\! x_{l,s\!}(t) \!\otimes\! x_{k\!-\!l,s\!}(t) \!+\!
\Gamma_{\!u} \!\!\left(\! x_{k \!-\! 1,s\!}(t) \!\otimes\! u(t) \!\right) \!+\! \overline{u}_{k}(t)
\label{app:1-12}
\end{eqnarray}
in which
\begin{eqnarray*}
& & \hspace*{-1.0cm} \overline{u}_{k}(t) \!\!=\!\!  \Gamma_{x} \!\!\sum_{l=1}^{k \!-\!1}\!\!\left[\!x_{l,t}(t) \!\otimes\! x_{k\!-\!l,t}(t) \!+\! x_{l,s}(t)\!\otimes\! x_{k\!-\!l,t}(t) +\right. \\
& & \hspace*{1.8cm}
\left. x_{l,t}(t) \!\otimes\! x_{k\!-\!l,s}(t)\!\right] \!+\!
\Gamma_{u}\!\left(x_{k \!-\! 1,t}(t)  \!\otimes\! u(t)\right)
\end{eqnarray*}

From Theorem \ref{theo:1}, it is clear that for every $k=1,2,\cdots$, in each term of $x_{k,t}(t)$, there is at least one mode of the NDS $\mathbf{\Sigma}_{p}$. It can therefore be declared directly from the definition of Kronecker matrix production that this claim is also valid for every term of the above vector $\overline{u}_{k}(t)$.

On the other hand, from Equations (\ref{eqn:theo:1-3}) and (\ref{app:1-5}), direct algebraic manipulations show that
\begin{eqnarray}
& & \hspace*{-0.8cm}
x_{l,s\!}(t) \!\otimes\! x_{k\!-\!l,s\!}(t) \!\!=\!\! \left(\sum_{i_{1}=1}^{m_{\xi}}\cdots\sum_{i_{l}=1}^{m_{\xi}}e^{\lambda_{s}(i_{h}|_{h=1}^{l})t}\psi_{s}(i_{h}|_{h=1}^{l})\right)
\!\!\otimes \nonumber \\
& & \hspace*{1.2cm}
\left(\sum_{i_{1}=1}^{m_{\xi}}\cdots\sum_{i_{k\!-\!l}=1}^{m_{\xi}}e^{\lambda_{s}(i_{h}|_{h=1}^{k\!-\!l})t}
\psi_{s}(i_{h}|_{h=1}^{k\!-\!l})\right) \nonumber \\
& & \hspace*{-0.4cm}
=\!\!\!\!   \sum_{i_{1}=1}^{m_{\xi}} \!\!\cdots \!\! \sum_{i_{k}=1}^{m_{\xi}} \!\!\!\! \left( \!\! e^{\lambda_{s\!}(i_{h\!}|_{h=1}^{k})t} \!\psi_{s\!}(i_{h}\!|_{h=1}^{l})
\!\otimes\! \psi_{s\!}(i_{h\!}|_{h=l\!+\!1}^{k})  \!\!\!\right)
\label{app:1-13}
\\
& & \hspace*{-0.8cm}
x_{k\!-\!1,s\!}(t) \!\otimes\! u(t) \!\!=\!\!\! \left(\! \sum_{i_{1}=1}^{m_{\xi}} \!\!\cdots\!\!  \sum_{i_{k\!-\!1}=1}^{m_{\xi}} \!\!\!\!e^{\lambda_{s}(i_{h}|_{h=1}^{k\!-\!1})t}\psi_{s}(i_{h}|_{h=1}^{k\!-\!1}) \!\!\right)
\!\!\otimes \nonumber \\
& & \hspace*{4.2cm}
\left(\sum_{i=1}^{m_{\xi}}e^{\lambda_{s,i}t} \psi_{u}(i) \right) \nonumber \\
& & \hspace*{-0.4cm}
=\!\!   \sum_{i_{1}=1}^{m_{\xi}} \!\!\cdots \!\! \sum_{i_{k}=1}^{m_{\xi}} \!\!\!\! \left( \!\! e^{\lambda_{s}(i_{h}|_{h=1}^{k})t} \!\!\psi_{s}(i_{h}|_{h=1}^{k\!-\!1})
\!\otimes\! \psi_{u}(i_{k})  \!\!\!\right)
\label{app:1-14}
\end{eqnarray}

Denote the Laplace transform of the VVF $\overline{u}_{k}(t)$ by $\overline{u}_{k}(s)$, and take Laplace transform for both sides of Equation (\ref{app:1-12}). Then the following equality can be directly obtained from Equations (\ref{app:1-13}) and (\ref{app:1-14}), in which $k = 2,3,\cdots$,
\begin{eqnarray}
& & \hspace*{-1.0cm} E \left[ s{x}_{k}(s) \!-\! {x}_{k}(0) \right]\!=\! Ax_{k}(s) \!+\! \overline{u}_{k}(s) \!+
\nonumber\\
& & \hspace*{-0.0cm} \!\!\!
\sum_{i_{1}=1}^{m_{\xi}} \!\!\cdots \!\! \sum_{i_{k}=1}^{m_{\xi}}
\frac{\Gamma_{\!x}\sum_{l=1}^{k \!-\!1}\!\psi_{s\!}(i_{h}\!|_{h=1}^{l})
\!\otimes\! \psi_{s\!}(i_{h\!}|_{h=l\!+\!1}^{k})}{s-\lambda_{s}(i_{h}|_{h=1}^{k})} \!+ \nonumber\\
& & \hspace*{-0.0cm} \!\!\!
\sum_{i_{1}=1}^{m_{\xi}} \!\!\cdots \!\! \sum_{i_{k}=1}^{m_{\xi}}
\frac{\Gamma_{\!u} \psi_{s}(i_{h}|_{h=1}^{k\!-\!1})
\!\otimes\! \psi_{u}(i_{k})}{s-\lambda_{s}(i_{h}|_{h=1}^{k})}
\label{app:1-15}
\end{eqnarray}
Hence
\begin{eqnarray}
& & \hspace*{-0.8cm} {x}_{k}(s)\!=\! (sE-A)^{-1}\left[ x_{k}(0) \!+\! \overline{u}_{k}(s) \right] \!+
\nonumber\\
& & \hspace*{-0.4cm} \!\!\!
(sE \!-\! A)^{\!-\!1} \!\! \sum_{i_{1}\!=\!1}^{m_{\xi}} \!\!\cdots \!\! \sum_{i_{k}\!=\!1}^{m_{\xi}} \!\!
\frac{\Gamma_{\!x} \!\sum_{l=1}^{k \!-\!1}\!\psi_{s\!}(i_{h}\!|_{h=1}^{l})
\!\otimes\! \psi_{s\!}(i_{h\!}|_{h=l\!+\!1}^{k})}{s-\lambda_{s}(i_{h}|_{h=1}^{k})} \!+ \nonumber\\
& & \hspace*{-0.4cm} \!\!\!
(sE-A)^{-1} \!\! \sum_{i_{1}=1}^{m_{\xi}} \!\!\cdots \!\! \sum_{i_{k}=1}^{m_{\xi}}
\frac{\Gamma_{\!u} \psi_{s}(i_{h}|_{h=1}^{k\!-\!1})
\!\otimes\! \psi_{u}(i_{k})}{s-\lambda_{s}(i_{h}|_{h=1}^{k})}
\label{app:1-16}
\end{eqnarray}

Recall that in each term of the VVF $\overline{u}_{k}(t)$, there is at least one mode of the NDS $\mathbf{\Sigma}_{p}$. Let
${x}_{k,s}(s)$ denote the Laplace transform of the steady-state part of ${x}_{k}(t)$, that is, the VVF ${x}_{k,s}(t)$. Then the above equation immediately leads to
\begin{eqnarray}
& & \hspace*{-0.8cm} {x}_{k,s}(s)\!=\!\!\! \sum_{i_{1}=1}^{m_{\xi}} \!\!\cdots \!\! \sum_{i_{k}=1}^{m_{\xi}}\!\!\!
\frac{\left(\lambda_{s}(i_{h}|_{h=1}^{k})E-A\right)^{\!\!-\!1}}{s-\lambda_{s}(i_{h}|_{h=1}^{k})} \!\!\left\{\!
\Gamma_{\!x} \!\sum_{l=1}^{k \!-\!1} \!\!\left[\! \psi_{s\!}(i_{h}\!|_{h=1}^{l}) \otimes \right.\right. \nonumber\\
& & \hspace*{0.4cm}
\left.\left. \psi_{s\!}(i_{h\!}|_{h=l\!+\!1}^{k})\right] + \Gamma_{\!u} \psi_{s}(i_{h}|_{h=1}^{k\!-\!1})
\!\otimes\! \psi_{u}(i_{k})\right\}
\label{app:1-17}
\end{eqnarray}
The proof can now be completed by taking inverse Laplace transform of both sides of the above equation, and recalling the definition of the scalar $\lambda_{s}(i_{h}|_{h=1}^{k})$. \hspace{\fill}$\Diamond$

\hspace*{-0.40cm}{\rm\bf Proof of Theorem \ref{theo:2}.} Note that the system matrices of both the NDS $\mathbf{\Sigma}_{p}$ and the assembly PSGS  $\mathbf{\Sigma}_{s}$ are real valued. It can therefore be declared that if they have a complex (generalized) eigenvalue, then its conjugate is also an (generalized) eigenvalue of them. Hence, when all the eigenvalues of the assembly PSGS  $\mathbf{\Sigma}_{s}$ are on the imaginary axis, they can be divided into several conjugate pairs\footnote[2]{The only exception is an eigenvalue located on the origin. In this case, with a little abuse of terminology, it can be regarded as a self-conjugate eigenvalue, and assign half of the associated quantities to each of them in the associated partial fraction decompositions. This treatment is helpful in avoiding awkward statements.}. This means that in the steady-state response $x_{k,s}(t)$ of the state vector $x_{k}(t)$ of the NDS $\mathbf{\Sigma}_{p}$, the pair $\left(\lambda_{s}(i_{h}|_{h=1}^{k}),\;\psi_{s}(i_{h}|_{h=1}^{k})\right)$ and the pair $\left(\overline{\lambda_{s}(i_{h}|_{h=1}^{k})},\; \overline{\psi_{s}(i_{h}|_{h=1}^{k})}\right)$ must exist simultaneously.

From the Euler formula, we have that
\begin{eqnarray}
& & \hspace*{-1.6cm}
cos\left(k\lambda_{s}^{[i]}(i_{h}|_{h=1}^{l}) T\right)=\frac{e^{k\lambda_{s}(i_{h}|_{h=1}^{l})T} + e^{k\overline{\lambda_{s}(i_{h}|_{h=1}^{l})}T}}{2}  \\
& & \hspace*{-1.6cm}
sin\left(k\lambda_{s}^{[i]}(i_{h}|_{h=1}^{l}) T\right)=\frac{e^{k\lambda_{s}(i_{h}|_{h=1}^{l})T} - e^{k\overline{\lambda_{s}(i_{h}|_{h=1}^{l})}T}}{2\mathbf{i}}
\end{eqnarray}
On the other hand, from the definitions of $\lambda_{s}(i_{h}|_{h=1}^{l})$ and $\lambda_{p,s}^{[l,q]}(i_{h}|_{h=1}^{l})$, it is clear that
\begin{eqnarray}
& & \hspace*{-1.6cm}
\pm\lambda_{s}(i_{h}|_{h=1}^{l}) + \lambda_{p,j_{1}} = \lambda_{p,j_{1}} \pm \sum_{h=1}^{l} \!\lambda_{s,i_{h}}  \\
& & \hspace*{-1.6cm}
\pm\lambda_{s}(i_{h}|_{h=1}^{l}) \!+\! \lambda_{p,j_{1}} \!+\!  \lambda_{p,s}^{[f,q]}(j_{h}|_{h=1}^{m}) \!=\! \sum_{h=2}^{q} \!\lambda_{p,j_{h}} \!\pm  \nonumber \\
& & \hspace*{1.6cm}  \sum_{h=1}^{l} \!\lambda_{s,i_{h}} \!+\! \sum_{j=q+1}^{f} \!\lambda_{s,j_{h}}  \\
& & \hspace*{-1.6cm}
\pm\lambda_{s}(i_{h}|_{h=1}^{l}) + \lambda_{s}(j_{h}|_{h=1}^{f}) = \pm\sum_{h=1}^{l} \!\lambda_{s,i_{h}} + \sum_{h=1}^{f} \!\lambda_{s,j_{h}}
\end{eqnarray}

Therefore, when the linear part of the NDS $\mathbf{\Sigma}_{p}$ is stable and the assembly PSGS  $\mathbf{\Sigma}_{s}$ has eigenvalues only on the imaginary axis, the real parts of both $\lambda_{s}(i_{h}|_{h=1}^{l}) + \lambda_{p,j_{1}}$ and $\lambda_{s}(i_{h}|_{h=1}^{l}) \!+\! \lambda_{p,j_{1}} \!+\!  \lambda_{p,s}^{[f,q]}(j_{h}|_{h=1}^{m})$ are always negative, while that of $\lambda_{s}(i_{h}|_{h=1}^{l}) + \lambda_{s}(j_{h}|_{h=1}^{f})$ is always equal to zero.

The proof can now be completed through a direct application of Theorem \ref{theo:1} and Lemma \ref{lemma:4}. \hspace{\fill}$\Diamond$

\hspace*{-0.40cm}{\rm\bf Proof of Theorem \ref{theo:3}.}
When the systems matrices $A(\theta)$, $B(\theta)$, $C(\theta)$ and $D(\theta)$ depend on the SIP vector $\theta$ through the LFT of Equation (\ref{plant-3}), it has been shown that the associated TFM also depends on this parameter vector through an LFT \cite{zly2024}. More specifically,
define TFMs $G_{zu}(s)$, $G_{zv}(s)$, $G_{yu}(s)$ and $G_{yv}(s)$ respectively as
\begin{eqnarray*}
	\left[\begin{array}{cc}
		G_{yu}(s) & G_{yv}(s)\\
		G_{zu}(s) & G_{zv}(s)
	\end{array}\right]
	& = &\!\!\!\!
	\left[\begin{array}{cc}
		D_{yu} & D_{yv}\\
		D_{zu} & D_{zv}
	\end{array}\right]
	+
	\left[\begin{array}{c}
		C_{yx}\\C_{zx}
	\end{array}\right]\times  \nonumber\\
& &	\!\!\!\!
	[s E -A_{xx}]^{-1}
	\left[\begin{array}{cc}
		B_{xu} & B_{xv}
	\end{array}\right]
\end{eqnarray*}
Then when the regularity assumption (Assumption 3) and the well-posedness assumption (Assumption 2) are satisfied, we have that
\begin{eqnarray}
H(s_{1}\!,\theta) \!\!\!\!&=&\!\!\!\! G_{\! yu}(s_{1}) + G_{\! yv}(s_{\!1})[I_{m_v} \!- \Theta(\theta) G_{\! zv}(s_{1})]^{-\!1} \!\!\times \nonumber\\
& & \hspace*{3.0cm} \Theta(\theta) G_{\! zu}(s_{1})
\label{eqn:lft-1}
\end{eqnarray}

From Equation (\ref{scm}), it is clear that $\Theta(\theta)$ depends affinely on each element of the NDS SIP vector $\theta$. It can therefore be declared that the TFM $H(s_{1},\theta)$ depends through an LFT on the NDS SIP vector $\theta$.

On the other hand, define TFMs $G_{\! xx}(s)$, $G_{\! zx}(s)$, $G_{\! xu}(s)$  and $G_{\! xv}(s)$ respectively as
\begin{eqnarray*}
& &
G_{\! xx}(s) =  [s E -A_{xx}]^{-1},  \hspace{0.50cm}
G_{\! zx}(s) = C_{zx} G_{\! xx}(s) \\
& &
G_{\! xu}(s) = G_{\! xx}(s) B_{xv}, \hspace{0.90cm}
G_{\! xv}(s) = G_{\! xx}(s)B_{xv}
\end{eqnarray*}
Then through similar arguments as those of \cite{Zhou2025} for the derivations of Equation (\ref{eqn:lft-1}), it can be shown that
\begin{eqnarray}
& & \hspace*{-0.8cm} [s E \!-\! A(\theta)]^{-1} \!=\! G_{\! xx}(s) + G_{\! xv}(s)[I_{m_v} \!- \Theta(\theta) G_{\! zv}(s)]^{-\!1} \!\!\times \nonumber\\
& & \hspace*{4.0cm} \Theta(\theta) G_{\! zx}(s)
\label{eqn:lft-2}  \\
& &  \hspace*{-0.8cm} [s E \!-\! A(\theta)]^{-1} \! B(\theta) \!=\! G_{\! xu}(s) + G_{\! xv}(s)[I_{m_v} \!- \Theta(\theta) \!\!\times \nonumber\\
& & \hspace*{3.0cm} G_{\! zv}(s)]^{-\!1} \Theta(\theta) G_{\! zu}(s)
\label{eqn:lft-3}
\end{eqnarray}
meaning that both $[s E \!-\! A(\theta)]^{-1}$ and $[s E \!-\! A(\theta)]^{-1}B(\theta)$ can also be expressed as an LFT of the NDS SIP vector $\theta$.

Note that both addition and multiplication of any two LFTs with compatible dimensions can still be expressed as an LFT \cite{zdg1996}. From Corollary \ref{coro:1} and Equations (\ref{eqn:lft-1})-(\ref{eqn:lft-3}), as well as the definition of the multi-dimensional/generalized TFM $H(s_{i}|_{i=1}^{k},\theta)$ with $k \geq 2$ that is given by Equation (\ref{eqn:tfm-2}), it is clear that for each $k \geq 2$ and any tuple $i_{h}|_{h=1}^{k}$, the multi-dimensional/generalized TFM $H(s_{i}|_{i=1}^{k},\theta)$ depends also through an LFT on the NDS SIP vector $\theta$. This completes the proof. \hspace{\fill}$\Diamond$

\end{document}